\documentclass[letterpaper]{ar-1col}
\usepackage{aas_macros}
\usepackage{rotating}%
\usepackage{natbib}%
\usepackage{amsfonts,amsmath,amssymb}
\usepackage[margin=1.5in]{geometry}
\jname{Xxxx. Xxx. Xxx. Xxx.}
\jvol{00}
\jyear{YYYY}
\doi{10.1146/((please add article doi))}

\usepackage{graphicx}
\usepackage{epstopdf}
\def\lesssim{\mathbin{\lower 3pt\hbox
    {$\rlap{\raise 5pt\hbox{$\char'074$}}\mathchar"7218$}}}   
\def\gtrsim{\mathbin{\lower 3pt\hbox
    {$\rlap{\raise 5pt\hbox{$\char'076$}}\mathchar"7218$}}}   

\newcommand{\fortyeight}{SAX~J1748.9$-$2021}
\newcommand{\twenty}{4U~1820$-$30}
\newcommand{\fortyfive}{EXO~1745$-$248}
\newcommand{\thirtyone}{KS~1731$-$260}
\newcommand{\twentyfour}{4U~1724$-$207}
\newcommand{\oeight}{4U~1608$-$52}
\newcommand{\twentyeight}{4U~1728$-$34}

\begin{document}

\markboth{Feryal \"Ozel and Paulo Freire}{Masses, Radii, and Equation of State of Neutron Stars}

\title{Masses, Radii, and Equation of State of Neutron Stars}

\author{Feryal \"Ozel$^1$ and Paulo Freire$^2$
\affil{$^1$Department of Astronomy, University of Arizona, Tucson, AZ 85721, USA; 
email: fozel@email.arizona.edu}
\affil{$^2$Max-Planck-Institut f\"ur Radioastronomie, Auf dem H\"ugel 69, D-53121, 
Bonn, Germany; email: pfreire@mpifr-bonn.mpg.de}}


\begin{abstract}
We summarize our current knowledge of neutron star masses and
radii. Recent instrumentation and computational advances have resulted
in a rapid increase in the discovery rate and precise timing of radio
pulsars in binaries in the last few years, leading to a large number
of mass measurements. These discoveries show that the neutron star
mass distribution is much wider than previously thought, with 3 known
pulsars now firmly in the $1.9-2.0\;M_{\odot}$ mass range.  For radii,
large, high quality datasets from X-ray satellites as well as
significant progress in theoretical modeling led to considerable
progress in the measurements, placing them in the $9.9-11.2$~km range
and shrinking their uncertainties due to a better understanding of the
sources of systematic errors. The combination of the massive neutron
star discoveries, the tighter radius measurements, and improved
laboratory constraints of the properties of dense matter has already
made a substantial impact on our understanding of the composition and
bulk properties of cold nuclear matter at densities higher than that
of the atomic nucleus, a major unsolved problem in modern physics.
\end{abstract}


\maketitle


\section{INTRODUCTION}

Our understanding of neutron stars has changed drastically since the
Annual Reviews article of \citet{wheeler1966}, when not a single
neutron star was known and the discussion consisted of an entirely
theoretical treatment of collapsed dense stars. The change has been
most rapid in the last decade, when the discovery and precise timing
observations of pulsars has proceeded at an ever accelerating rate,
the new generation X-ray and $\gamma$-ray telescopes have provided
high quality, large datasets, and a large body of theoretical work on
neutron star emission properties and spacetimes have enabled
significant recent developments in our ability to model these extreme
objects and interpret the observations in solid frameworks.

We now know precise masses for $\sim 35$ neutron stars spanning the
range from 1.17 to 2.0 $M_\odot$ and can pin down the radii of more
than a dozen to the 9.9$-$11.2~km range. The combination of the heaviest
known neutron star mass with the existing radius measurements already
places significant constraints on the cold dense matter equation of
state up to densities that are 8 times the nuclear density.

Certain characteristics of neutron stars have been known for a while.
Shortly after the discovery of pulsars \citep{hewish1968}, it became
apparent that they are the observational manifestation of extremely
compact stars made primarily of neutrons. It was also evident that,
while the neutron degeneracy pressure played a role in supporting
these compact objects against gravitational collapse, it was not
sufficient to hold up a star beyond $0.7 M_\odot$
\citep{oppenheimer1939,tolman1939}, and that repulsive nuclear forces
were necessary in shaping their structure \citep{harrison1965}.
Earlier calculations of the interiors treated nucleons with
approximate potentials and predicted masses in the $0.5-3 M_\odot$
range and radii between 7 and 20 km. Stars of such extreme compactness
have central densities that are 5-10 times the nuclear saturation
density $\rho_{\rm sat} = 2.8 \times 10^{14}$~g~cm$^{-3}$.

While the nuclear models improved over the years, the major advances
have come from astrophysical observations and models because
terrestrial experiments do not approach densities similar to that of
neutron star cores (see section 4.2). Theoretical work is guided by
the well developed effective field theory approaches and the rapidly
developing QCD calculations but there remain very large uncertainties
in our understanding of the actual compositions of the cores. Open
questions include the composition of matter at high densities, such as
the density at which matter needs to be described by quarks and no
longer by nucleons, as well as the presence of strangeness or boson
condensates in such matter; the isospin symmetry energy that defines
the energy difference between normal and neutron-rich matter and,
ultimately, the ability to calculate matter in $\beta$-equilibrium
from first principles; and the role and reliability of calculations of
n-body interactions. Despite the large number of possibilities, each
theoretical model, or its resulting equation of state, can be mapped
into a mass-radius relation by solving the general relativistic
structure equations. This, in turn, allows measurements of the neutron
star masses and radii to place strong constraints on the properties
and interactions of cold, ultradense matter, as we will discuss in
this review. These same observations also provide excellent tests of
general relativity in the strong-field regime, using pulsars as well
as neutron star surfaces as testbeds.

The recent series of investigations of neutron star masses and radii
are motivated not only by the aim of probing physics in new regimes,
but also because of the forefront role neutron stars play in many
astrophysical phenomena. Neutron stars are associated with numerous
explosive, transient, and non-electromagnetic events, and neutron star
properties play a role in shaping each one of them. The primary
sources of gravitational waves that are expected to be detected with
the gravitational wave detector Advanced LIGO are the neutron
star-neutron star and neutron star-black hole mergers. These systems
are also thought to be the central engines of short gamma-ray bursts
(see \citealt{nakar2007} for a review). The dynamics of the mergers,
the gravitational waveforms in late stages of coalescence, the
lifetime of merger disks and the resulting black hole formation
timescales, as well as any accompanying bursts of neutrino, gamma-ray,
and/or optical emission depends sensitively on the neutron-star mass,
radius, and the equation of state
\citep{baiotti2008,metzger2010,hotokezaka2011,bauswein2012,lackey2012,kumar2015}.

From a stellar lifecycle and source population point of view, studies
of neutron stars prove to be equally important. Different equations of
state allow different maximum masses for neutron stars and, thus,
determine the dividing line between neutron stars and black
holes. This has a direct impact on the outcomes of supernova
explosions and the nature of compact remnants, as well as on the
numbers of neutron stars and black holes in the Galaxy. This, in turn,
affects the number of observable compact object binaries, their
properties, and possible merger/gravitational-wave event rates. In
fact, the supernova mechanism itself is affected by the equation of
state of ultradense matter. The hot dense matter EoS, an extension of
the cold one probed by neutron stars, determines explosion conditions
(e.g., \citealt{janka2007}) and is crucial for understanding core
collapse supernova explosions and the associated phenomena including
mass loss, r-process nucleosynthesis, gravitational wave and neutrino
emission.

The tremendous advance in the measurements of neutron star masses,
radii, and the dense matter equation of state have come from a diverse
array of techniques, applied to many different populations of neutron
stars, and enabled by observations in all wavelengths from the radio
to gamma rays.  For example, while the precise mass measurements have
largely resulted from radio observations of pulsars, the radius
measurements have been almost exclusively been a result of X-ray
observations of neutron stars in low-mass X-ray binaries.  In this
review, we will focus on the latest mass and radius measurements of
neutron stars and their implications for the dense matter equation of
state, as well as a number of questions that remain open.

\section{NEUTRON STAR MASS MEASUREMENTS}

The vast majority of the precise mass measurements of neutron stars
have been performed using radio observations of rotation-powered
pulsars. Currently more than 2500 pulsars are known in the Galaxy
\citep{manchester2005}, most of which are detectable as radio pulsars,
but also some observed in X-rays and an increasingly large number
detected in gamma
rays\footnote{https://confluence.slac.stanford.edu/display/GLAMCOG/Public+List+of+LAT-Detected+Gamma-Ray+Pulsars}
\citep{abdo2013,caraveo2014}.

About 90\% of radio pulsars are isolated. Their masses cannot be
measured, since all of the current methods rely on precise tracking of
the orbital motions through the arrival times of the observed
pulsations.  The remaining 250 pulsars are located in binary systems
(three of these are in multiple component systems). We will now focus
exclusively on these.

Most of the pulsars in binaries are ``recycled'': at some point during
the evolution of their companion there was mass transfer from the
companion to the neutron star. The mass transfer can, in principle,
increase the pulsar mass significantly, but its clearest effect is the
spin-up of the pulsar (to spin frequencies as high as 716~Hz, see
\citealt{hessels2006}) and -- by mechanisms that are poorly understood
-- a sharp reduction in the pulsar's magnetic field, to values smaller
than $10^{11}$~G.  This process produces a wide variety of binary
pulsars (for a recent review see, e.g., \citealt{lorimer2008}).  Those
of greatest interest to us are the systems where both components are
compact: the double neutron star systems (DNSs) and the millisecond
pulsar - white dwarf (MSP-WD) systems.

\subsection{Timing binary pulsars}

The extraordinary long-term rotational stability of recycled pulsars
and their fast rotation makes them uniquely useful for timing. If a
recycled pulsar is in a binary system (as it often happens), then we
can use this precise timing to measure its orbital motion with
astounding precision.

In Newtonian gravity, the part of the orbital motion we can observe
(generally along the line of sight only) can be described by five
Keplerian parameters: the binary period $P_b$, the orbital
eccentricity $e$, the projection of the pulsar's semi-major axis
$a_{\rm PSR}$ onto the observer's line of sight $x_{\rm PSR} = a_{\rm
  PSR} \sin i / c$ (where $i$ is the angle between the orbital angular
momentum vector and the line of sight and $c$ is the speed of light),
the time of periastron $T_0$ and the longitude of periastron $\omega$.

For each new pulsar, these parameters are determined by fitting a
radial velocity model to the observed spin periods, which are Doppler
shifted by the orbital motion of the pulsar. This is similar to what
can be done in spectroscopic systems. The unique feature of pulsars is
that, by determining the correct rotational phase count, one can
directly {\em range} the pulsar relative to the center of mass of the
system.  Measuring a pulse time of arrival with a precision of 1
$\mu$s translates into a ranging accuracy of 300~m per measurement,
which is much smaller than the astronomical sizes of the orbits. This
provides Keplerian parameters many orders of magnitude more precise
than those derived from Doppler shift measurements with the same data
and is the reason why pulsars are so useful for such a wide variety of
purposes - including uniquely precise mass measurements.

The binary mass function is then determined from the Keplerian orbital
parameters
\begin{equation}
\label{eq:f}
f = \frac{(M_{\rm c} \sin i)^3}{M_{\rm T}^2} = \frac{4 \pi^2}{T_{\odot}}
\frac{x_{\rm PSR}^3}{{P_b}^2},  
\end{equation}
where $T_{\odot} \equiv G M_{\odot}/c^3 = 4.925490947\,\mu$s is the
mass of the Sun in time units\footnote{The product $G M_{\odot}$,
  where $G$ is Newton's gravitational constant and $M_{\odot}$ is the
  solar mass is very precisely known, despite the fact that neither
  $G$ nor $M_{\odot}$ are individually known to better than one part
  in $10^3$.} , and $M_{\rm PSR}$,
$M_{\rm c}$ and $M_{\rm T} = M_{\rm PSR} + M_{\rm c}$ are the pulsar,
companion and total binary masses, respectively, all expressed in solar
masses. This single equation has three unknowns ($i, M_{\rm PSR}$ and
$M_{\rm c}$); therefore, in the absence of further information, the
individual masses $M_{\rm psr}$ and $M_{\rm c}$ cannot be determined
despite the very high precision in the determination of the Keplerian
parameters.

If the projected semi-major axis of the companion's orbit $x_{\rm c}$
can be measured, either via timing, if the companion is a pulsar, or
via phase-resolved optical spectroscopy, if the companion is a white
dwarf or main-sequence star, one extra equation is obtained:
\begin{equation}
\label{eq:q}
q \equiv \frac{M_{\rm psr}}{M_{\rm c}} = \frac{x_{\rm c}}{x_{\rm psr}},
\end{equation}
which is valid, to first Post-Newtonian order, in all conservative
theories of gravity. The number of binary systems where we can measure
$q$ is small, but even for those systems, measuring $f$ and $q$ is not
enough to determine the masses. This can be addressed in a few systems
since the spectrum of a white dwarf companion allows, in some cases, a
determination of its mass, $M_{\rm WD}$ (see section 2.3.3).

Because of the small sizes of neutron stars and their neutron star or
white dwarf companions compared to their orbits, both components can
be treated as point masses. In this case, the ``classical'' tidal and
rotational effects that complicate the motion of other astrophysical
objects and hide the (generally smaller) relativistic effects are
simply absent, i.e., such systems are clean gravitational
laboratories. In such a system, a pulsar with good timing precision
will allow the detection and precise measurement of these small
relativistic effects. These effects can be parameterized by
``post-Keplerian'' (PK) parameters, which, assuming that general
relativity (GR) is the correct theory of gravity, and to leading
post-Newtonian order, are related to three of the Keplerian parameters
($P_b$, $x$ and $e$) and the system masses.  For the five PK
parameters that have been measured in the context of pulsar timing,
these expressions are as follows (as in \citealt{stairs2003} and
references therein):
\begin{enumerate}
\item The rate of advance of periastron $\dot{\omega}$, typically
  measurable in eccentric orbits.  This is analogous to the anomalous
  perihelion advance in Mercury's orbit:
\begin{equation}
\label{eq:omegadot}
\dot{\omega} =  3 \left( \frac{P_{b}}{2\pi} \right)^{-5/3}
\left( T_{\odot} M_{\rm T} \right)^{2/3}\left(1-e^2 \right)^{-1}.
\end{equation}
\item The ``Einstein delay'', $\gamma$.
In GR, this is due in equal parts to the variations of the
gravitational redshift and of special relativistic time dilation in
an eccentric orbit:
\begin{equation}
\label{eq:gamma}
\gamma = e \left( \frac{P_{b}}{2\pi} \right)^{1/3}
T_{\odot}^{2/3} M_{\rm T}^{-4/3}
M_{\rm c} \left( M_{\rm PSR} + 2 M_{\rm c} \right).
\end{equation}
\item The orbital period decay $\dot{P}_{b}$ due to the loss of
  orbital energy of the system caused by the emission of gravitational
  waves (this typically requires timing observations that span many
  years),
\begin{equation}
\label{eq:pbdot}
\dot{P}_b = -\frac{192\pi}{5} \left( \frac{P_{b}}{2 \pi T_{\odot}} \right)^{-5/3}  
\left( 1+ \frac{73}{24}e^2 + \frac{37}{96} e^4 \right) \times \\
(1-e^2)^{-7/2}\, M_{\rm PSR} M_{\rm c}\,M_{\rm T}^{-1/3}.
\end{equation}
\item The range $r$ and the shape $s$ of Shapiro delay, which is a
  delay in the reception of the radio pulses at the Earth that is due
  to the propagation of the radio signal in the curved space-time near
  the companion star
\begin{eqnarray}
r & = & T_{\odot} M_{\rm c},\\
s = \sin i & = & x_{\rm PSR} \left(\frac{P_{b}}{2\pi}\right)^{-2/3}
T_{\odot}^{-1/3} M_{\rm T}^{2/3}\, M_{\rm c}^{-1}.
\label{eq:shapiro}
\end{eqnarray}
This effect is more easily detected in edge-on orbits.
\end{enumerate}

\begin{figure}
\includegraphics[height=3in]{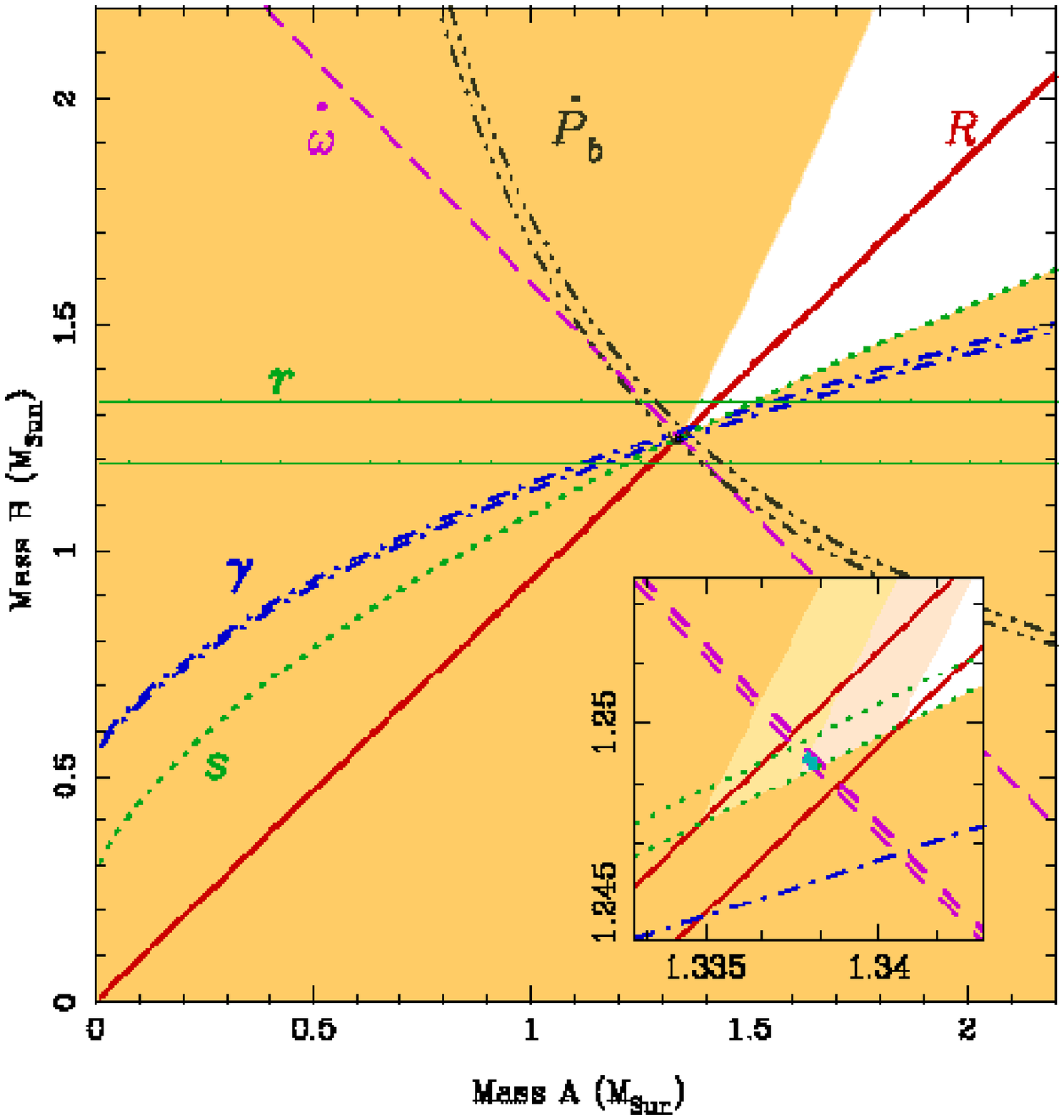}
\caption{Mass-mass diagram for the J0737$-$3039 system.  The yellow
  areas are excluded by the mass functions of both pulsars,
  PSRs~J0737$-$3039A and B.  The lines represent the regions allowed
  by eqs.~\ref{eq:q} to \ref{eq:shapiro} for the measured PK
  parameters.  All lines meet at a single point in the diagram, i.e.,
  GR passes the tests posed by these four distinct constraints. For
  our purposes, this plot is very important since it indicates that we
  get consistent mass measurements using any pair of PK
  parameters. From Kramer et al. 2006.}
\label{fig:0737}
\end{figure}

We must here emphasize that, like the Keplerian parameters, these are
parameters that appear in the equations that describe the motion and
timing of binary pulsars, such as those presented in
\citet{damour1985,damour1986}. They are measured by by timing a
particular system and fitting the pulse arrival times. It is only when
we adopt a specific theory of gravity that we can relate them to
physical parameters of the components of the binary - in the case of
GR, and to leading PN order, their masses only (eqs.~\ref{eq:omegadot}
- \ref{eq:shapiro}). These constraints are displayed graphically in
Fig.~\ref{fig:0737}. Note that there is nothing unique about the
particular parameterization of \citet{damour1985}. As an example,
instead of $r$ and $s$, we can describe Shapiro delay with two less
correlated PK parameters $\varsigma$ and $h_3$ \citep{freire2010}.

\subsection{Double neutron star mass measurements}

The first system for which reliable mass measurements were made was
PSR~B1913+16, the first binary pulsar ever discovered
\citep{hulse1975}. The compact orbit of this double neutron star
system ($P_b = 0.32299$ days) and its high orbital eccentricity ($e =
0.6171334(5)$) allowed a precise measurement of the PK parameter
$\dot{\omega}$. Subsequently, \citet{taylor1979} were able to measure
two more PK parameters, $\gamma$ and $\dot{P}_b$. The individual
masses were determined from Eqs.~\ref{eq:omegadot} and \ref{eq:gamma}
and the observed orbital decay was then used as a successful test of
GR. The GR prediction for $\dot{P}_b$ caused by gravitational wave
emission (Eq.~\ref{eq:pbdot}) for the derived masses was consistent
observed rate (see \citealt{damour2015} for a review).


\begin{table}
\tabcolsep14.25pt
\caption{Masses of Double Neutron Star Systems and Non-recycled Pulsars}
\label{tab1}
\begin{center}
\begin{tabular}{@{}llllcc@{}}
\hline
System     & $M_{\rm T}$   & $M_{\rm PSR}$ & $M_{\rm c}$   & Mass    & Ref. \\
           & ($M_{\odot}$) & ($M_{\odot}$) & ($M_{\odot}$) & const.  &      \\
\hline
\multicolumn{6}{c}{Systems with well-measured component masses}\\
\hline
J0453+1559       & 2.734(4)    & 1.559(5)   & 1.174(4)   & $\dot{\omega},h_3$    & 1,2 \\
J0737$-$3039     & 2.58708(16) & 1.3381(7)  & 1.2489(7) y& $\dot{\omega},q$      & 3 \\
B1534+12         & 2.678463(8) & 1.3330(4)  & 1.3455(4)  & $\dot{\omega},\gamma$ & 4 \\
J1756$-$2251     & 2.56999(6)  & 1.341(7)   & 1.230(7)   & $\dot{\omega},\gamma$ & 5 \\ 
J1906+0746       & 2.6134(3)   & 1.291(11) y & 1.322(11) ? & $\dot{\omega},\gamma$ & 6 \\
B1913+16         & 2.828378(7) & 1.4398(2)  & 1.3886(2)  & $\dot{\omega},\gamma$ & 7 \\
B2127+11C g      & 2.71279(13) & 1.358(10)  & 1.354(10)  & $\dot{\omega},\gamma$ & 8 \\  
\hline
\multicolumn{6}{c}{Systems with total binary mass measurement only}\\
\hline
J1518+4904       & 2.7183(7)   &   $<$1.768    & $>$0.950   & $\dot{\omega}$     & 9 \\
J1811$-$1736     & 2.57(10)    &   $<$1.64     & $>$0.93    & $\dot{\omega}$     & 10 \\
J1829+2456       & 2.59(2)     &   $<$ 1.34    & $>$1.26    & $\dot{\omega}$     & 11 \\
J1930$-$1852     & 2.59(4)     &   $<$ 1.32    & $>$1.30    & $\dot{\omega}$     & 12 \\
\hline
\multicolumn{6}{c}{Non-recycled pulsars with massive WD companions}\\
\hline
J1141$-$6545     & 2.2892(3)   & 1.27(1) y   & 1.01(1)   & $\dot{\omega},\gamma$     &14,15 \\
B2303+46         & 2.64(5)     & 1.24-1.44 y & 1.4-1.2   & $\dot{\omega},M_{\rm WD}$  &15,16 \\
\hline 
\end{tabular}
\end{center}
\footnotesize{Notes: The systems indicated with a ``g'' are located in globular clusters. A question mark 
indicates that the NS nature of the companion is not firmly established. The mass measurements for 
neutron stars detected as normal (non-recycled) radio pulsars are indicated with the letter
  ``y''.  References are to the latest mass measurements:
  1.\ \citet{deneva2013} 2.\ \citet{martinez2015}
  3.\ \citet{kramer2006} 4.\ \citet{fonseca2014}
  5.\ \citet{ferdman2014} 6.\ \citet{leeuwen2015} 
  7.\ \citet{weisberg2010} 8.\ \citet{jacoby2006} 
  9.\ \citet{janssen2008} 10.\ \citet{corongiu2007} 
  11.\ \citet{champion2005} 12.\ \citet{swiggum2015} 
  13.\ \citet{bhat2008} 14.\ \citet{antoniadis2011} 
  15.\ \citet{thorsett1999} 16.\ \citet{kerkwijk1999}}
\end{table}

A number of other double neutron systems (DNSs) have since been
discovered.  The eleven systems for which mass constraints are known
are listed in the top two groups in Table~\ref{tab1}, together with
the latest references. These systems are relatively rare, about 5\% of
the currently known population of binary pulsars. We list the total
mass $M_{\rm T}$ of each binary since, generally, that is much better
known than the individual masses of the components. This happens
because the orbits of all double neutron stars are eccentric, so that
$\dot{\omega}$ (and $M_{\rm T}$) can be determined with high
precision.  In binaries where additional PK parameters are measured,
we also list the individual NS masses and specify the parameters that
were measured to derive the masses (for all these systems,
Eq.~\ref{eq:f} can be used to determine the orbital inclination
$i$). For one of these systems -- J1906+0746 -- it is possible for the
companion to be a massive WD instead. PSR~B2127+11C is located in the
globular cluster M15 and is very likely the result of an exchange
interaction \citep{prince1991,lynch2012}.

PSR~B1913+16 is no longer unique in providing a stringent test of
general relativity. The orbital decay due to the emission of
gravitational waves has been measured for {\em five} other DNS systems
(J0737$-$3039A/B, B1534+12, J1756$-$2251, J1906+0746 and B2127+11C),
and, as we will discuss below, a few millisecond pulsar-white dwarf
systems as well.

We also include in Table~\ref{tab1} two radio pulsars that have
precise mass measurements even though they are not members of DNS
systems. Their slow spins and higher magnetic fields indicate that
they have not been recycled, making them resemble DNSs rather than the
millisecond pulsars with white dwarf companions that we discuss in
Section~2.3.

\subsubsection{The double pulsar}

Among the DNS population, one system in particular stands out:
J0737$-$3039. This system is a unique laboratory for gravitational
physics because it combines an unusual number of desirable features:
{\it (i)}\ With an orbital period of 2h 27m, it is by far the most
compact DNS known. Combined with the moderate orbital eccentricity of
the system, this implies that the PK parameters $\dot{\omega}$,
$\gamma$ and $\dot{P}_b$ are relatively large and will be easy to
measure, promising one precise test of GR. {\it (ii)}\ The system has
a very high orbital inclination, $\sim 89^\circ$. This allows for a
very precise measurement of the two Shapiro delay parameters, $r$ and
$s$, thus providing two more independent tests of GR. {\it (iii)}\ The
second neutron star in the system is also an active radio pulsar,
which is detectable at least some of the time \citep{lyne2004}. This
makes it the only currently known {\em double pulsar} system and
allows a precise measurement of the mass ratio, $q$, which ``frees''
one of the PK parameters from the role of mass determination and,
therefore, allows one more test of GR.

The four tests in this system provided by eqs.~\ref{eq:omegadot} -
\ref{eq:shapiro} are remarkably stringent, but despite that, GR has
passed all of them \citep{kramer2006}. This can be seen in
Fig.~\ref{fig:0737}, where all the regions compatible with the known
mass constraints are consistent with the same pair of masses.  Such
tests of gravity theories are not the topic of this review --- we
refer interested readers to \citet{wex2014} for a comprehensive
discussion.  Nevertheless, systems like J0737$-$3039 and
Fig~\ref{fig:0737} are important for NS mass measurements because they
demonstrate experimentally that any pair of mass constraints chosen
from $q$ and the PK parameters will generally yield consistent NS
masses.

\subsection{Millisecond pulsars}

\begin{table}
\caption{Masses of Millisecond Pulsars}
\label{tab2}
\begin{center}
\begin{tabular}{@{}llllcc@{}}
\hline
System     & $M_{\rm T}$   & $M_{\rm PSR}$ & $M_{\rm c}$   & Mass    & Ref. \\
           & ($M_{\odot}$) & ($M_{\odot}$) & ($M_{\odot}$) & const.  &      \\
\hline
\multicolumn{6}{c}{MSPs with WD companions and low-eccentricity orbits}\\
\hline
J0348+0432       &             & 2.01(4)              & 0.172(3)       & $q, M_{\rm WD}$ & \citet{antoniadis2013} \\
J0437$-$4715     &             & 1.44(7)              & 0.224(7)       & $r,s$          & \citet{reardon2016} \\
J0621+1002       &   2.32(8)   & $1.53^{+0.10}_{-0.20}$ & $0.76^{+0.28}_{-0.07}$ & $\dot{\omega}, s$ & \citet{kasian2012}  \\
J0751+1807       &             & 1.72(7)              & 0.13(2)        & $s, \dot{P}_b$ & \citet{desvignes2016} \\
J1012+5307       &             & 1.83(11)             & 0.16(2)        & $q, M_{\rm WD}$ & \citet{antoniadis2016} \\
J1614$-$2230     &             & 1.928(17)            & 0.500(6)       & $r,s$          & \citet{fonseca2016} \\ 
J1713+0747       &	       & 1.31(11)             &	0.286(12)      & $r,s$          & \citet{zhu2015} \\
J1738+0333       &             & $1.47^{+0.07}_{-0.06}$ & $0.181^{+0.007}_{-0.005}$ & $q, M_{\rm WD}$ & \citet{antoniadis2012} \\
J1802$-$2124     &             & 1.24(11)             &	0.78(4)        & $r,s$         & \citet{ferdman2010} \\
J1807$-$2500B    & 2.57190(73) & 1.3655(21)           & 1.2064(20)(?)  & $\dot{\omega},h_3$ & \citet{lynch2012} \\
B1855+09         &             & $1.58^{+10}_{-13}$    & $0.267^{+0.010}_{-0.014}$  & $r,s$ & \citet{splaver2004} \\ 
J1909$-$3744     &             & 1.47(3)              &	 0.2067(19)    & $r,s$         & \citet{reardon2016} \\
J2222$-$0137     &             & 1.20(14)             & 1.05(6)        & $r,s$         & \citet{kaplan2014} \\
\hline
\multicolumn{6}{c}{MSPs with eccentric orbits and triples}\\
\hline
J0337+1715       &             & 1.4378(13)           & 0.19751(15)  & $i, q$        & \citet{ransom2014,kaplan2014a} \\
                 &             &                      & 0.4101(3)    &               &        \\
J1903+0327       &  2.697(29)  &  1.667(21)           & 1.029(8)     & $\dot{\omega}, h_3$  &  \citet{freire2011}\\
J1946+3417       &  2.097(28)  &  1.832(28)           & 0.2659(30)   & $\dot{\omega}, h_3$  & \citet{barr2016}\\
J2234+0611       &  1.668(6)   &  1.393(13)           & 0.276(9)     & $\dot{\omega}, h_3$  & \citet{stovall2015}\\
\hline
\multicolumn{6}{c}{MSPs in globular clusters}\\
\hline
J0024$-$7204H   &  1.61(4)      &   $< 1.52$    &     $> 0.164$  & $\dot{\omega}$ & \citet{freire2003}  \\
J0514$-$4002A   &  2.453(14)   &   $< 1.50$    &     $> 0.96$   & $\dot{\omega}$ & \citet{freire2007} \\
B1516+02B       &  2.29(17)     &   $< 2.52$    &     $> 0.13$   & $\dot{\omega}$ & \citet{freire2008a} \\
J1748$-$2021B   &  2.92(20)     &   $< 3.24$    &     $> 0.11$   &  $\dot{\omega}$ & \citet{freire2008} \\
J1748$-$2446I   &  2.17(2)      &   $< 1.96$    &     $> 0.24$   & $\dot{\omega}$ & \citet{ransom2005} \\
J1748$-$2446J   &  2.20(4)      &   $< 1.96$    &     $> 0.38$   & $\dot{\omega}$ & \citet{ransom2005} \\
J1750$-$37A     & 1.97(15)      &   $<1.65$     &     $> 0.53$   &  $\dot{\omega}$ & \citet{freire2008} \\
B1802$-$07      & 1.62(7)       &   $< 1.7$     &     $> 0.23$   &  $\dot{\omega}$ & \citet{thorsett1999} \\
J1824$-$2452C   & 1.616(7)      &   $< 1.35$    &     $> 0.26$   &  $\dot{\omega}$ & \citet{begin2006} \\               
J1910-5958A     &               &   $1.3(2)$    &     0.180(18)  &  $q, M_{\rm WD}$ & \citet{bassa2006,cocozza2006} \\
\hline 
\end{tabular}
\end{center}
\footnotesize{Notes: J1807$-$2500B is located in the globular cluster
  NGC6544. A question mark indicates that the nature of the companion
  is uncertain. The total mass is indicated only when it is known more
  precisely than the masses of the components.  References are to the
  latest mass measurements. }
\end{table}


The term ``millisecond pulsars'' (MSPs) refers to pulsars with spin
periods in the range $1.39\, < \, P \, < \, 20$~ms and
$\dot{P} < 10^{-19}$.
These systems have been heavily recycled (and circularized)
by a long-lived accretion phase in a low-mass X-ray binary (LMXB).
The first MSP by this definition, B1937+21, was discovered at the
Arecibo telescope in 1982 \citep{backer1982}. (At a 642~Hz spin
frequency, this was the fastest known MSP for the subsequent 24
years.) A very large number of MSPs have been discovered since and,
although $\sim$20\% are isolated, most of the remaining objects have
white dwarf companions and very small orbital eccentricities.

The fast (and very stable) rotation of MSPs makes their timing
significantly more precise than for the pulsars in DNSs.  However,
their low orbital eccentricities pose a problem for mass determination
(and tests of GR), since, in these cases, the PK parameters
$\dot{\omega}$ and especially $\gamma$ cannot be measured
accurately. Furthermore, most systems have orbital periods larger than
0.5~d, which makes a detection of the very small predicted
$\dot{P}_b$ nearly impossible. These difficulties made the measurement
of MSP masses a slow and difficult process. Four strategies have been
followed over the last two decades, which have recently started
producing a number of precise mass measurements.

\subsubsection{Pulsars in globular clusters}
One way of getting around the issue of orbital circularity is to find
MSPs in globular clusters (GCs). Because of the much higher stellar
density, there are substantial gravitational interactions with nearby
stars that, with time, can make an initially circular orbit acquire a
substantial eccentricity. This allows the measurement of
$\dot{\omega}$ and, if there are no classical contributions to this
effect, the total mass of the system $M_{\rm T}$ for quite a few
systems (see Table~\ref{tab2}).

The first binary mass precise enough to be useful was that of
PSR~J0024$-$7204H, a 3.21 ms pulsar in a 2.35-day, slightly eccentric
($e = 0.070560(3)$) orbit around a low-mass WD in the globular cluster
47~Tucanae, which yielded $M_{\rm T} \, = \, 1.61 \, \pm \, 0.04 \,
M_{\odot}$ \citep{freire2003}. No further PK parameters were
measurable; nevertheless, one can infer from combining Eqs.~\ref{eq:f}
and \ref{eq:omegadot} (and the mathematical fact $\sin i \leq 1$) that
$M_{\rm PSR} \,< \,1.52 \,M_{\odot}$.  This was one of the earlier
indications that the long recycling episodes that spin up MSPs do not
necessarily make them very massive.

Most eccentric systems discovered since in GCs share the same
characteristic: we can measure $\dot{\omega}$ but no other mass
constraints.  The reason is that the most eccentric systems tend to be
the ones with wider orbits, which are easier to perturb by nearby
stars. For these relatively wide orbits, it is very difficult to
measure $\gamma$ and $\dot{P}_b$.

The two exceptions are two of the exchange encounter systems B2127+11C
and J1807$-$2500B listed in Tables~\ref{tab1} and \ref{tab2}. Both are
located in clusters with collapsed cores, where exchange encounters
are more likely \citep{verbunt2014}. The orbital period of B2127+11C
is 8.0 hr and its orbital eccentricity is 0.681386(16), which allows
measurements of $\dot{\omega}$, $\gamma$ and even $\dot{P}_b$
\citep{jacoby2006}.  The last system, J1807$-$2500B, is particularly
important for this discussion: it is a true MSP with a spin period is
4.18 ms, but despite the long accretion episode, its current mass is
only 1.3655(21) $M_{\odot}$. More than any other system, this MSP
demonstrates that full recycling can be achieved without a large mass
increment.

Nevertheless, as the number of $M_{\rm T}$ measurements accumulated,
it became apparent that some systems, such as PSRs~J1748$-$2446I and J
in Terzan 5 \citep{ransom2005}, PSR~J1748$-$2021B in NGC~6440
\citep{freire2008} and PSR~B1516+02B in M5 \citep{freire2008a} are
quite massive but have very low mass functions. This is only likely to
happen if the pulsars and not the companions in these systems are
massive. Indeed, by 2008, there was already strong evidence that a
fair fraction of MSPs must have masses close to 2 $M_{\odot}$
\citep{freire2008a}. This statistical results was confirmed by the
later MSP mass measurements, described below.

\subsubsection{Shapiro delay measurements}
For the vast majority of pulsar-WD systems, the only measurable PK
parameters are those related to the Shapiro delay. However, the
fastest-spinning MSPs tend to have very low-mass helium white dwarf
companions, which reduces the amplitude of the Shapiro delay.
Therefore, in order to measure the masses in these systems, both a
very good timing precision and a high orbital inclination are
required.

The first MSP-WD system where the Shapiro delay was clearly detected
was PSR~B1855+09 \citep{ryba1991}. However, until recently, the timing
precision in this system was not enough for a precise determination of
the pulsar mass. This changed in 2003 with the discovery of
PSR~J1909$-$3744, a heavily recycled MSP with $P = 2.9$ ms and a light
helium white dwarf companion \citep{jacoby2003}.  This system combines
very high timing accuracy with a very high orbital inclination, which
allowed, in 2005, the first precise measurement of the mass of a MSP:
$1.438\, \pm \, 0.024 M_{\odot}$ \citep{jacoby2005}.  

Since then, the number of MSPs discovered has been increasing
significantly (see section~\ref{sec:MSPnumbers}), and for that reason
the number of precise Shapiro delay mass measurements has been
accumulating, particularly for pulsars with more massive carbon-oxygen
(CO) or oxygen-neon-magnesium (O-Ne-Mg) white dwarf companions, such
as PSR~J1802$-$2124 \citep{ferdman2010}, or more recently
PSR~J2222$-$0137 \citep{kaplan2014}.  The larger companion masses make
the Shapiro delay easier to measure; this is why a disproportionate
number of MSPs with well-measured masses have such massive companions.
This does not apply as a rule to the MSPs themselves: the latter
systems highlight the fact that many NSs are born with masses as low
as $\sim 1.2\, M_{\odot}$.

Of these MSPs with massive WD companions, the measurement that had by
far the greatest impact was PSR~J1614$-$2230 \citep{demorest2010}.
The originally reported mass of the pulsar was 1.97(4) $M_{\odot}$,
exceptional and unexpected among the MSPs with massive WD
companions. This mass measurement had implications for the NS mass
distribution as well as for the NS equation of state (EOSs), which
will be discussed in later sections. Note that this measurement has
since been slightly improved, which revised the mass to
$M=1.928(7)\;M_\odot$ \citep{arzoumanian2015}, still consistent with
the earlier value.


\subsubsection{Spectroscopic mass measurements}

For some of the MSPs, particularly those with short orbital periods,
the companion is optically bright. This has allowed reliable mass
determinations of both components by studying the Balmer lines
produced by hydrogen in the white dwarf atmosphere.

First, phase-resolved spectoscopy can measure how these lines shift
with orbital phase, providing a measurement of the projected orbital
velocity of the white dwarf. Comparing this to the orbital velocity of
the pulsar measured in the timing ($v_{\rm PSR} = 2 \pi x_{\rm PSR} /
P_{b}$) we can immediately determine the mass ratio, $q \equiv M_{\rm
  PSR} / M_{\rm WD} = v_{\rm WD}/v_{\rm PSR}$.

Second, the width of the Balmer lines is related to the local
gravitational acceleration, $g = G M_{\rm WD} / R_{\rm WD}^2$.  Using
model relations between $M_{\rm WD}$ and $R_{\rm WD}$ for white dwarfs
(see, e.g., the references in \citealt{antoniadis2012}), both quantities
can be determined from $g$. After this step $M_{\rm PSR}$ can be
obtained trivially from $q M_{\rm WD}$.

The first such measurement was made for PSR~J1012+5307
\citep{callanan1998}, later followed by the mass measurement for
J1910$-$5958A, a MSP-WD system in NGC~6752
\citep{bassa2006,cocozza2006}.  To date, the two most important such
measurements have been those of J1738+0333 \citep{antoniadis2012} and
J0348+0432 \citep{lynch2013,antoniadis2013}.  For the former $M_{\rm
  PSR} = 1.47^{+0.07}_{-0.06} M_{\odot}$ and $M_{\rm WD} =
0.181^{+0.007}_{-0.005} M_{\odot}$; this measurement was important
because the relatively short orbital period (8.5 hr) allowed a
measurement of $\dot{P}_b$ \citep{freire2012}, which agrees with the
GR prediction for these masses.  Given the asymmetry in the binding
energy of this system, this measurement introduced the most stringent
constraints ever for several families of gravity theories
\citep{freire2012}, superseding the previous test from the pulsar - WD
system PSR~J1141$-$6545 \citep{bhat2008}.

For PSR~J0348+0432, a pulsar with a WD companion with an orbital
period of only 2hr 27m, Antoniadis et al. (2013) obtained $M_{\rm PSR}
= 2.01(4) M_{\odot}$ and $M_{\rm WD} = 0.172(3) M_{\odot}$. This
confirmed, using a different method, that neutron stars could reach
masses of the order of 2 $M_{\odot}$. Furthermore, the very short
orbital period has allowed a measurement of $\dot{P}_b$
\citep{antoniadis2013}, which again agrees with the GR prediction and
has already placed significant constraints on non-linear phenomena
predicted by some alternative theories of gravity for these massive
NSs, such as spontaneous scalarization \citep{damour1996}. 

\subsubsection{Triples, disrupted triples, and other monsters}

The theoretical expectation that the recycling pathway for fast MSPs
result in highly circular orbits was corroborated by the fact that all
fully recycled MSPs outside of globular clusters observed until 2008
had orbital eccentricities between $10^{-7}$ and $10^{-3}$
\citep{manchester2005}. In 2008, the situation changed with the
discovery of PSR~J1903$+$0327 \citep{champion2008}.  This
fast-spinning MSP ($P = 2.15$\,ms) is in binary system with a
$\sim1\,M_{\odot}$ main-sequence star companion in a 95-day, eccentric
orbit ($e=0.44$). This system is thought to have originated from a
hierarchical triple that became chaotically unstable, as suggested by
observational data \citep{freire2011} and numerical simulations
\citep{portegies-zwart2011}. This finding suggested triple stellar
systems containing MSPs might still be found in the Galaxy, which was
later confirmed with the discovery of PSR~J0337+1715, the first MSP in
a stellar triple system with two white dwarf companions
\citep{ransom2014}.  Note that the mass indicated in Table~\ref{tab2}
for PSR~J0337+1715 is based on these triple interactions.

Soon after the discovery of PSR~J1903+0327, \citet{bailes2010}
reported an ``anomalous'' orbital eccentricity of 0.027 for
PSR~J1618$-$3919, a 12-ms pulsar first reported in \citet{edwards2001}
that is in a 22-day orbit with a low-mass white dwarf companion. More
recently, \citet{deneva2013}, \citet{barr2013}, \citet{knispel2015},
and \citet{camilo2015} reported the discovery of four more unusual
binary MSP systems with eccentric orbits: PSRs~J2234$+$06,
J1946$+$3417, J1950+2414, and J0955$-$6150.  These four systems are
fully recycled with spin periods between 2 and 4.4 \,ms, orbital
periods from 22 to 32\,d and median companion masses $M_{\rm c}
\approx 0.25\,M_\odot$; i.e., apart from their large orbital
eccentricities ($0.07\, < \, e\, < \,0.14$) all parameters are
compatible with the canonical recycling formation channel leading to a
MSP with a white dwarf companion. For a discussion on their unusual
formation, see Knispel et al. (2015).


The reason why these systems are mentioned at length here is that for
nearly all of them, precise mass measurements have been, or will soon
be, made.  For PSR~J1903+0327, PSR~J1946+3417, and J2234+0611 precise
mass measurements have been obtained from the combination of
$\dot{\omega}$ and the Shapiro delay parameter $h_3$
(\citep{freire2011,barr2016,stovall2015}. Note that for
PSR~J1903+0327, the uncertainty quoted in Table~\ref{tab2} corresponds
to 99.7\% confidence level. The same will certainly be possible for
PSR~J1950+2414.
For the triple system, even more precise masses for the MSP and two
white dwarf companions could be obtained from the 3-body effects in
the system \citep{ransom2014}.

The masses in these systems are quite varied.  While the MSP in the
triple system J0337+1715 and J2234+0611 have masses of 1.4378(13)
$M_\odot$ and 1.393(13) $M_\odot$, respectively, PSR~J1946+3417 is in
the massive neutron star group, with $M_{\rm PSR} \, = \, 1.912(10) \;
M_\odot$ (Barr et al. 2015, in preparation).  PSR~J1903+0327 sits in
the middle, with $M_{\rm PSR} \, = \, 1.667(21) \; M_\odot$ (99.7\%
C. L.)\footnote{The error bars for this measurement are not Gaussian,
  therefore the 68\% error bar is not 1/3 of this uncertainty. See the
  original reference for a discussion.}

\subsection{Neutron Stars in X-ray Binaries}

Mass measurements have also been carried out for neutron stars with
high and low stellar mass companions using the observations of these
binaries in X-ray and optical wavelengths. We discuss the methods and
results for these two classes below.

\subsubsection{Neutron Stars with High-Mass Companions}

For mass measurements of neutron stars with high-mass companions,
eclipsing X-ray pulsars, in which the companions block the X-rats from
the pulsar for part of the orbit, form the primary targets.  In these
systems, X-ray observations of the pulsar yield the orbital period of
the binary $P_b$, the eccentricity of the orbit $e$, longitude of
periastron $\omega_0$, the semi-major axis of the neutron star's orbit
$a_X\sin i$, and the semi-duration of the eclipse $\theta_e$.  In
addition, optical observations of the companion star give its velocity
amplitude $K_{\rm opt}$, its projected rotational velocity $v_{\rm
  rot} \sin i$, and the amplitude of ellipsoidal variations $A$.  From
these observables, it is possible to solve for the fundamental
parameters of the binary, which include the mass of the neutron star
$M_{\rm NS}$, the mass, radius and rotational angular velocity of the
companion, $M_{\rm opt}$, $R_{\rm opt}$, $\Omega_{\rm opt}$, and the
inclination angle of the binary $i$.

These measurements are typically less precise than those obtained from
radio pulsar timing, and several sources of systematic uncertainty in
this technique have been discussed in \citet{ozel2012b} and
\citet{falanga2015}. The ten neutron stars whose masses have
been estimated in this way are listed in Table~\ref{tab3}.

\begin{table}
\caption{Masses of Neutron Stars in High-Mass and Low-Mass X-ray Binaries}
\label{tab3}
\begin{center}
\begin{tabular}{@{}lccc@{}}
\hline
System & $M_{\rm NS}$ & Error & References \\
& ($M_{\odot}$) & ($M_{\odot}$) & \\
\hline
\multicolumn{4}{c}{Neutron Stars in High-Mass X-ray Binaries}\\
\hline
LMC X$-$4&	    $1.57$&         $0.11$&    1 \\	
Cen X$-$3&	    $1.57$&         $0.16$&    1 \\
4U~ 1538$-$522&     $1.02$&         $0.17$&    1 \\
SMC X$-$1& 	    $1.21$&         $0.12$&    1 \\	
SAX J1802.7$-$2017& $1.57$&         $0.25$&    1 \\	
XTE J1855$-$026&    $1.41$&         $0.24$&    1 \\	
Vela X$-$1&	    $2.12$&         $0.16$&    1 \\
EXO 1722$-$363&     $1.91$&         $0.45$&    1 \\
OAO 1657$-$415&     $1.74$&         $0.30$&    1 \\
Her X$-$1& 	    $1.07$&         $0.36$&    1 \\	
\hline
\multicolumn{4}{c}{Neutron Stars in Low-Mass X-ray Binaries}\\
\hline
\oeight &       $1.57$&      $^{+0.30}_{-0.29}$& 2 \\
\twentyfour &   $1.81$&      $^{+0.25}_{-0.37}$& 2  \\
\thirtyone &    $1.61$&      $^{+0.35}_{-0.37}$& 2  \\
\fortyfive &    $1.65$&      $^{+0.21}_{-0.31}$& 2 \\
\fortyeight &   $1.81$&      $^{+0.25}_{-0.37}$& 2  \\
\twenty &       $1.77$&      $^{+0.25}_{-0.28}$& 2  \\
Cyg X$-$2&      $1.90$&      $^{+0.22}_{-0.35}$& 3 \\
\hline
\end{tabular}
\end{center}
\footnotesize{Notes. 1. See \citet{falanga2015}, \citet{ozel2012b},
  \citet{rawls2011}, and references therein. We exclude 4U~1700$-$377,
  for which there is no evidence that it is a neutron star. 2. See
  \citet{ozel2015} for the latest constraints. 3. \citet{orosz1999}.}
\end{table}

\subsubsection{Neutron Stars in Low-Mass X-ray Binaries}

Spectral studies of thermonuclear X-ray bursts from neutron stars
provide simultaneous measurements of their masses and radii. These
mass constraints are less precise than those obtained from pulsar
timing but still provide an estimate of neutron star masses in a
different population. The details of the methods will be discussed in
Section 3.1 and the mass measurements for the six sources in this
category are listed in Table~\ref{tab3}. This table also includes the
low-mass X-ray binary Cyg X$-$2, for which the mass of the neutron
star was inferred using optical observations of its companion.

\begin{figure}
\includegraphics[height=5in]{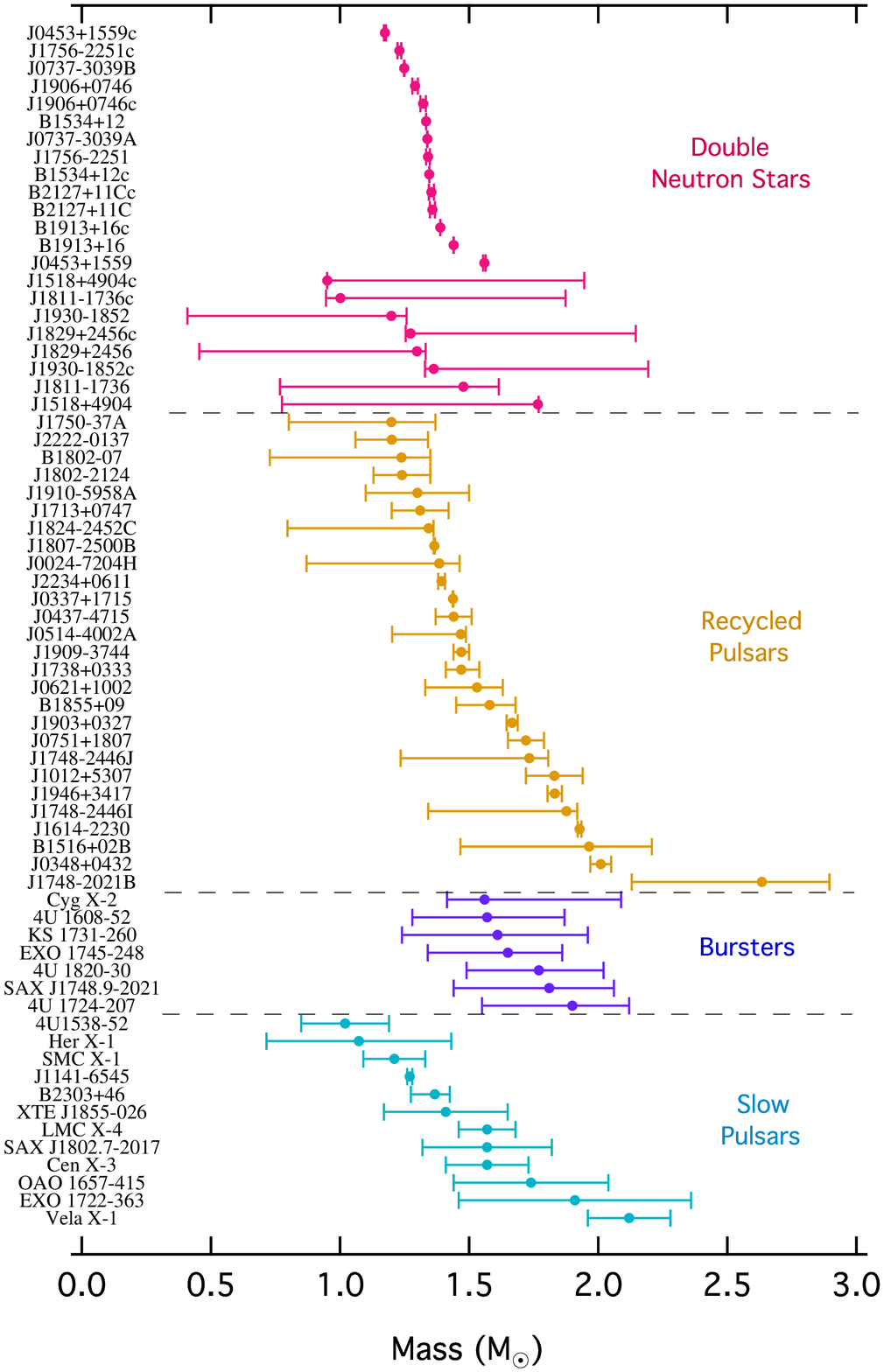}
\caption{The most recent measurement of neutron star masses. Double
  neutron stars (magenta), recycled pulsars (gold), bursters (purple),
  and slow pulsars (cyan) are included.}
\label{fig_masses}
\end{figure}

\subsection{Neutron Star Mass Distribution}

\begin{figure}
\includegraphics[height=3.0in]{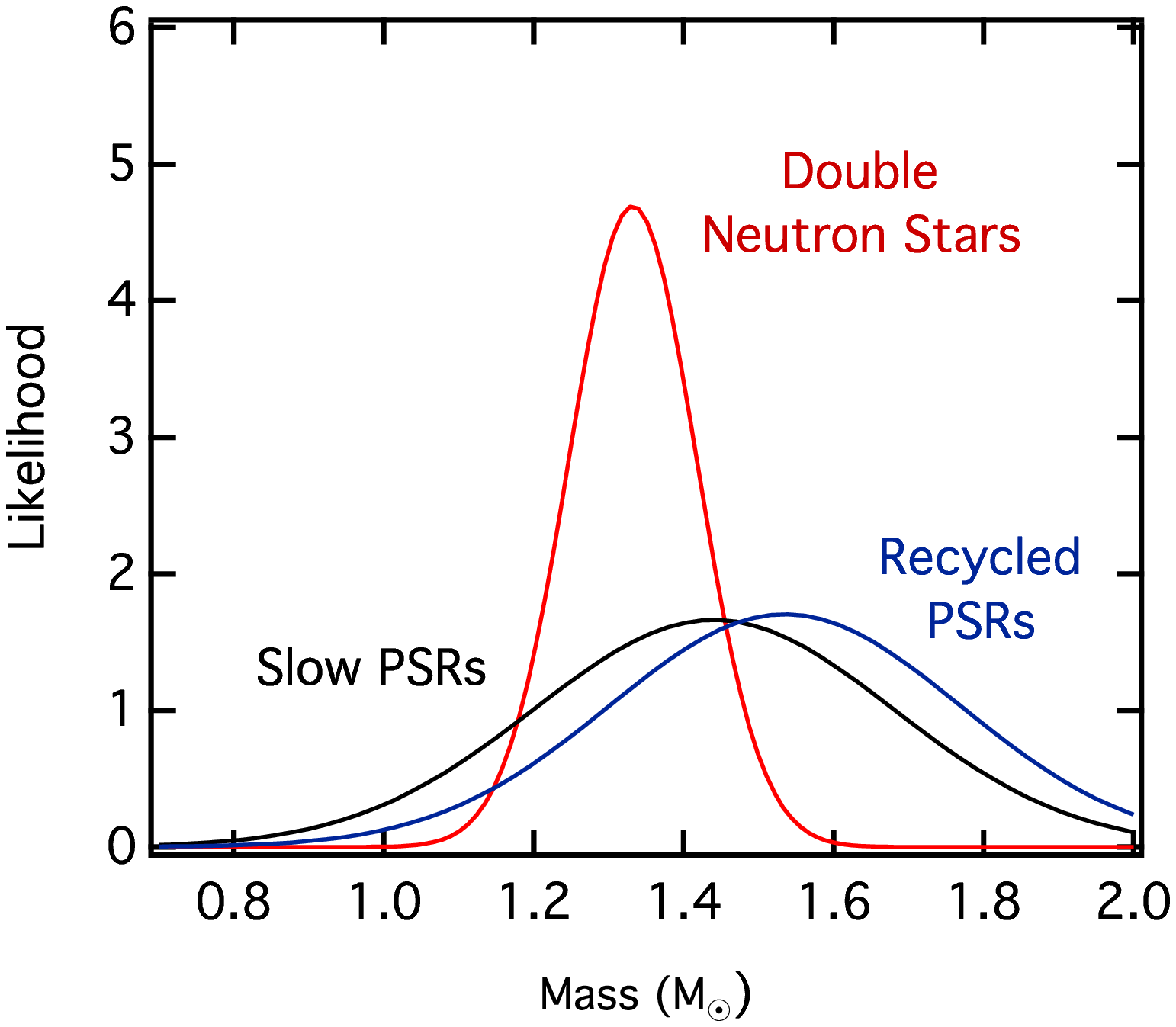}
\caption{The inferred mass distributions for the different populations
  of neutron stars.}\label{fig_mass_distr}
\end{figure}

We show in Figure~\ref{fig_masses} the combined neutron star mass
measurements in all of the above categories. It is already clear from
this figure that current measurements show a broad range of neutron
star masses, from $\approx 1.1-2 \; M_\odot$. The differences between
the neutron star masses in different categories are also evident.  To
study and characterize the mass distributions of these different
classes in more detail, it is possible to use Bayesian statistical
techniques on the currently available measurements. In particular, the
three different categories of sources, namely, the DNSs, the slow
pulsars (i.e., the small spin period pulsars and neutron stars with
high mass companions, which are likely to be near their birth masses)
and the recycled pulsars (which include all MSPs and the accreting
neutron stars with low-mass companions) can each be modeled with
Gaussian functions with a mean of $M_0$ and a dispersion $\sigma$
\begin{equation}
P(M_{\rm NS}; M_0,\sigma) = \frac{1}{\sqrt{2 \pi \sigma^2}} 
\exp\left[-\frac{(M_{\rm NS}-M_0)^2}{2 \sigma^2}\right].
\label{eq:gauss}
\end{equation} 
Several studies have employed Bayesian techniques to measure the most
likely values of the mean and dispersion for these systems
\citep{ozel2012b,kiziltan2013}. Fig.~\ref{fig_mass_distr} shows the
inferred mass distributions for these different categories of neutron
stars. The most likely values of the parameters for these
distributions are: $M_0=1.33 \; M_\odot$ and $\sigma=0.09 \; M_\odot$
for the DNSs, $M_0=1.54 \; M_\odot $ and $\sigma=0.23 \; M_\odot$ for
the recycled neutron stars, and $M_0=1.49 \; M_\odot $ and
$\sigma=0.19 \; M_\odot$ for the slow pulsars. A recent study also
raised the possibility of two peaks within the recycled millisecond
pulsar population, with the first peak at $M=1.388 \; M_\odot$ and a
dispersion $\sigma = 0.058 \;M_\odot$ and a second peak appearing at
$M=1.814 \; M_\odot$ with a dispersion of $\sigma=0.152 \; M_\odot$
\citep{antoniadis2016}.

Among these inferred distributions, the narrowness of the DNS
distribution stands out. While clearly not representative of neutron
stars as a whole, as it was once thought \citep{thorsett1999}, it
probably points to a particular evolutionary mechanism that keeps the
masses of neutron stars in these systems in a narrow range.  Recent
discoveries, such as the DNS J0453+1559 \citep{deneva2013}, indicate
that the range of masses in double neutron-star systems may also be
wider than previously believed: the recycled pulsar has a mass of
1.559(5) $M_{\odot}$, the heaviest known in any DNS
\citep{martinez2015}, while the companion has a mass of 1.174(4)
$M_{\odot}$, the smallest precisely measured mass for any NS (We infer
that the companion is a NS from the orbital eccentricity of the
system, $e = 0.11251837(5)$, which would not arise if it had slowly
evolved to a massive white dwarf star).

\subsection{Maximum Mass of Neutron Stars}

Finding the maximum mass of neutron stars is of particular interest in
mass measurements because of its direct implications for the neutron
star equation of state and neutron star evolution. The largest neutron
star mass can rule out the equations of state that have maximum masses
that fall below this value. The current record holder on this front is
J0348+0432 with a mass of $2.01\pm 0.04 M_\odot$
\citep{antoniadis2013}.

There are also some studies of a particular class of millisecond
pulsars called black widows (and their cousins redbacks) that have
suggested higher neutron star masses (e.g.,
\citealt{van-kerkwijk2011}). These millisecond pulsars irradiate and
ablate their very low-mass companions.  While the pulsar timing
provides the Keplerian parameters for the orbit, all other information
about the masses in these systems is obtained from the modeling of the
optical lightcurves (to determine orbital inclination) and the
spectroscopy (to measure the mass ratio) of the companion
star. Unfortunately, there are many difficulties in obtaining accurate
measurements from these ablated companions. Even when using a model of
an irradiated companion, the short timescale variability, the unevenly
heated surface, and the observed asymmetry in the lightcurves
significantly hinder the orbital inclination measurements. Similarly,
the difference between the center of light of the irradiated companion
and its center of mass as well as the departure of the spectral
features from assumptions of thermodynamic equilibrium introduce large
uncertainties in the inferred mass ratios \citep{roberts2015}. These
results are tantalizing and hint at the possibility of even more
massive neutron stars than J0348+0432, but because of the systematics,
we find that they are not yet as robust as the results from radio
timing.

\section{RADIUS MEASUREMENTS}

Neutron star radius measurements have progressed significantly in the
past decade and a number of different techniques have been
employed. Nearly all of the methods that are currently used rely on
the detection of thermal emission from the surface of the star either
to measure its apparent angular size or to detect the effects of the
neutron-star spacetime on this emission to extract the radius
information. The approaches can broadly be divided into spectroscopic
and timing measurements. In addition, there are proposed methods for
determining neutron star radii which show promise for the future but
will not be covered in this review, because there are currently
limited data, or because they provide only broad limits:
\begin{enumerate}
\item Quasi-periodic oscillations observed from accreting neutron stars (e.g., \citealt{miller1998})
\item Neutron star cooling (e.g., \citealt{page2006}) 
\item Pulsar glitches, which help constrain the relative thicknesses of neutron 
star crust vs. its core (e.g., \citealt{link1999})
\item Asteroseismology (e.g., \citealt{watts2015})
\end{enumerate}
 
\subsection{Spectroscopic measurements}

Much like measuring the radii of normal stars, spectroscopic
measurements of neutron star radii rely on determining their angular
sizes by measuring a thermal flux $F$, modeling the spectrum to
determine the effective temperature $T_{\rm eff}$, and combining this
with a distance measurement $D$ to obtain observed (or apparent) radii
\begin{equation} 
\frac{R_{\rm obs}}{D} = \left( \frac{F_{\rm bol}}{\sigma_{\rm B}
  T_{\rm eff}}\right)^{1/2},
\end{equation}
where $\sigma_{\rm B}$ is the Stefan-Boltzmann constant.  Unlike
stars, however, there are several complications that come into play in
this process. First, owing to their extreme compactness, neutron stars
gravitationally lense their own surface emission. This introduces
mass-dependent corrections to the observed angular
sizes \citep{pechenick1983,psaltis2000}, i.e., the physical radii are
related to the observed radii via 
\begin{equation}
R_{\rm obs} = \left(1-\frac{2GM}{Rc^2}\right)^{-1/2} R. 
\label{eq_rapp}
\end{equation}
(This equation holds as long as the radius is larger than the photon
orbit $3GM/c^2$. If it is smaller, then the geometric radius observed
at infinity would be equal to $\sqrt{27} GM/c^2$, independent of the
stellar radius.) The situation is even more complex in the case of
neutron stars that spin moderately fast, because their spacetime can
no longer be described by a Schwarzschild metric and there are further
spin-dependent corrections to the observed angular sizes
\citep{baubock2015}. Second, the surface emission may be contaminated
by non-thermal emission from an isolated neutron star's magnetosphere
or by the disk surrounding an accreting one. Third, the magnetic field
a neutron star possesses can be strong enough to cause temperature
non-uniformity on the surface, which complicates the inference of the
true stellar radius. Fourth, distances to neutron stars are, in
general, more difficult to measure than to normal stars.

Some of these challenges are overcome by detailed theoretical modeling
of emission from neutron stars in general relativity. Others are
mitigated or eliminated by careful source selection (e.g., sources
with low magnetic fields, with low accretion luminosities, or located
in globular clusters with known distances) and by combining multiple
observational phenomena to break the inherent degeneracies between
neutron star mass and radius and to reduce the measurement
uncertainties.  With these criteria as a guide, thermal emission from
low-mass X-ray binaries in quiescence and during X-ray bursts have
been the focus of recent work on neutron star radii.

\subsubsection{Quiescent Low-mass X-ray Binaries}

The first group of sources on which radius measurement efforts have
focused on are the accreting neutron stars in low-mass X-ray binaries
when they are in quiescence (qLMXBs).

In quiescence, accretion ceases or continues at a very low level. This
allows observations of the thermal emission from the stellar surface,
which is powered by the reradiation of the heat stored in the deep
crust during the accretion phases \citep{brown1998}. Because of the
very short ($\sim$mins) settling time of heavier elements in the
surface layers of neutron stars \citep{brown2002}, the atmospheres of
neutron stars in quiescence are expected to be composed of
hydrogen. (There are cases where the companion is hydrogen poor and a
helium atmosphere may be appropriate; see below). The observed spectra
of qLMXBs are indeed quasi-thermal, with a weak power-law component at
high energies that is likely due to residual accretion.

A number of qLMXBs in globular clusters has been observed with {\it
  Chandra} and XMM-{\it Newton}. Their luminosities in quiescence are
of the order of $10^{32-33}$~erg~s$^{-1}$, making them very faint
objects. In addition, the globular cluster environments entail crowded
fields. Because of this, the high angular resolution and low
background of the modern X-ray instruments were crucial for obtaining
spectroscopic constraints of their apparent angular sizes (e.g.,
\citealt{rutledge2001,heinke2006,webb2007}). 

To make quantitative measurements from the observed spectra, radiative
equilibrium models of neutron star atmospheres
\citep{zavlin1996,heinke2006} have been developed for a variety of
surface compositions, surface gravities and temperatures (for a recent
review, see \citealt{ozel2013}). Owing to the low magnetic field
strengths of neutron stars in low-mass X-ray binaries
(\citealt{psaltis1999} and references therein), and the expectation of
hydrogen atmospheres, unmagnetized hydrogen models have primarily been
used in the spectral analyses to obtain observed angular sizes.

\begin{figure}
\begin{minipage}{2.9in}
\includegraphics[scale=0.37]{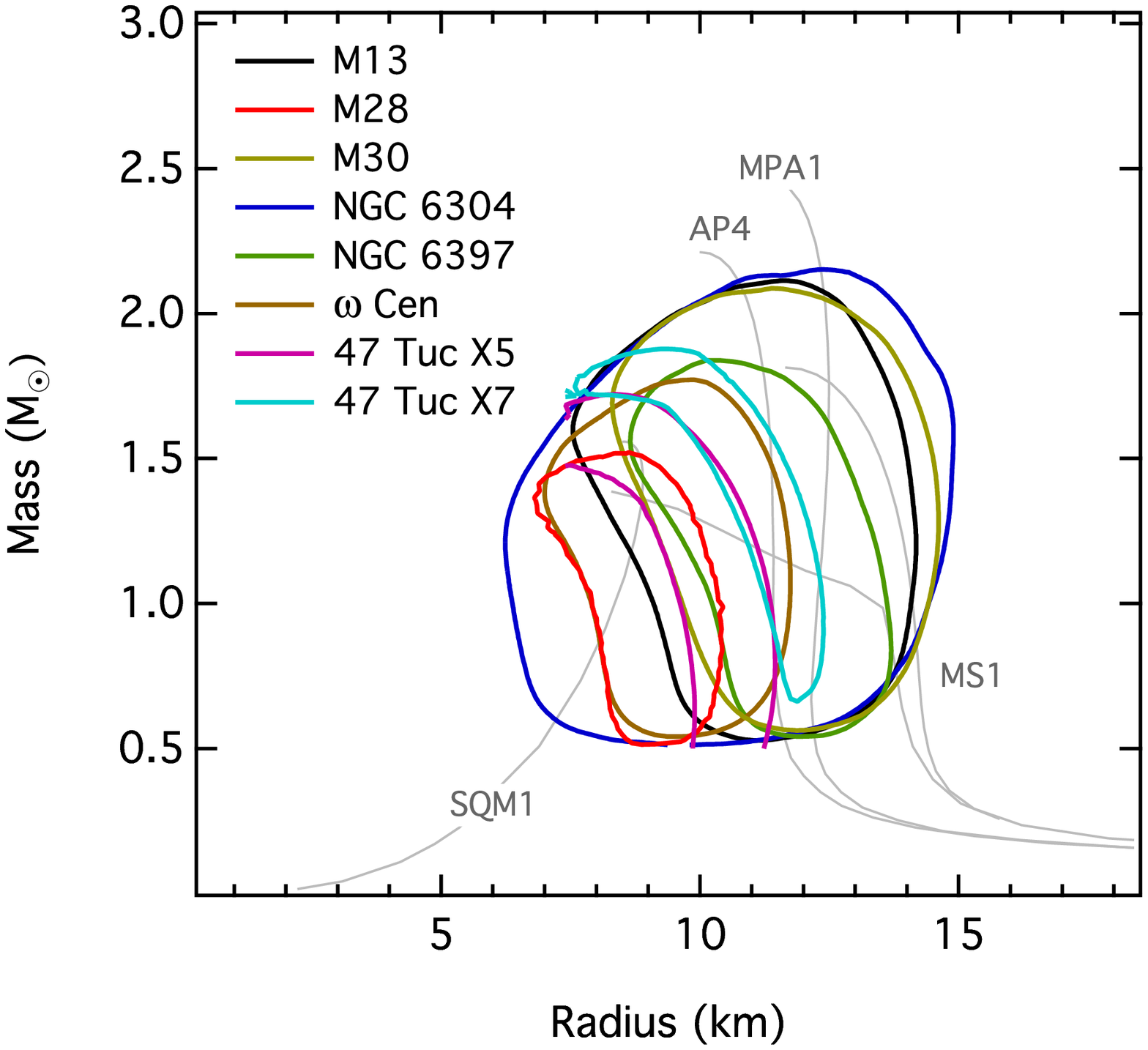}
\end{minipage}
\hfill
\begin{minipage}{2.9in}
\includegraphics[scale=0.37]{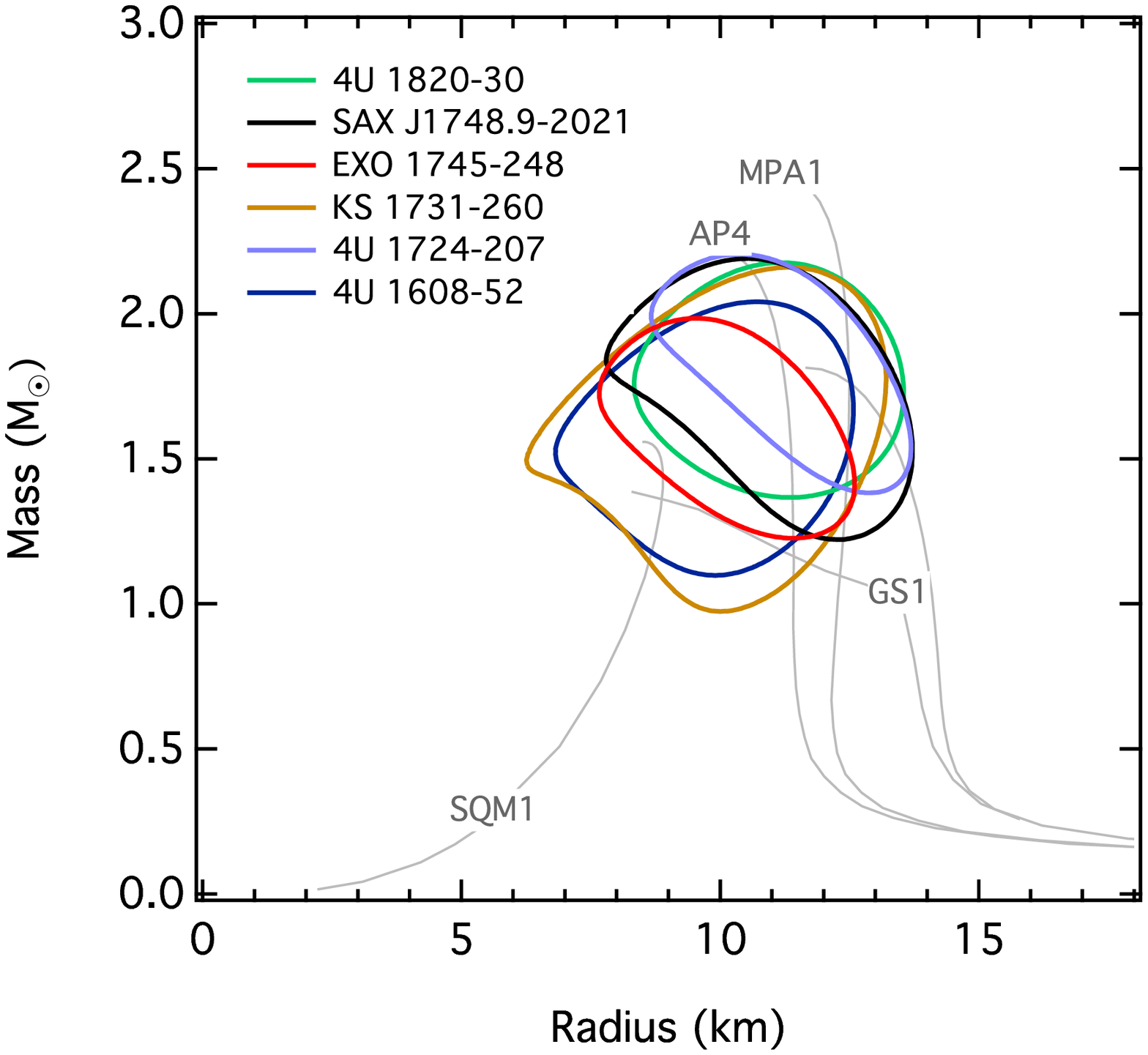}
\end{minipage}
\caption{The combined constraints at the 68\% confidence level over
  the neutron star mass and radius obtained from (Left) all neutron
  stars in low-mass X-ray binaries during quiescence (Right) all
  neutron stars with thermonuclear bursts. The light grey lines show
  mass-relations corresponding to a few representative equations of
  state (see Section 4.1 and Fig.~\ref{fig_EoS} for detailed
  descriptions.)}
\label{fig_spec_radii}
\end{figure}

To date, reliable radius constraints have been obtained for eight
qLMXBs located in the globular clusters M13, M28, M30, $\omega$~Cen,
NGC~6304, NGC~6397, and two in 47~Tuc
\citep{heinke2006,webb2007,guillot2011,bogdanov2016}. The observed
(apparent) angular sizes obtained for these sources have been combined
with globular cluster distances
\citep{harris1996,guillot2014,heinke2014} to measure their apparent
radii. Under somewhat different assumptions, there have been several
combined analyses of these measurements
\citep{guillot2013,guillot2014,lattimer2014,ozel2015}. The most recent
results are displayed as correlated contours on the neutron-star
mass-radius diagram\footnote{The full mass-radius likelihoods and
  tabular data for these sources can be found at
  http://xtreme.as.arizona.edu/NeutronStars.} (see
Fig.~\ref{fig_spec_radii}).

Several sources of systematic uncertainties that can affect the radius
measurements have been studied, which we discuss in some detail below. 

\paragraph{Atmospheric Composition}
The majority of qLMXBs for which optical spectra have been obtained
show evidence for H$\alpha$ emission \citep{heinke2014}, indicating a
hydrogen rich companion. Although none of these spectra have been
obtained for globular cluster qLMXBs, assuming that sources in
globular clusters have similar companions to those in the field led to
the use of hydrogen atmospheres when modeling quiescent spectra. There
is one source among the six that have been analyzed in detail, for
which there is evidence to the contrary. There is only an upper limit
on the H$\alpha$ emission from the qLMXB in NGC~6397 using HST
observations \citep{heinke2014}. Because of this, this source has been
modeled with a helium atmosphere and the corresponding results are
displayed in Fig.~\ref{fig_spec_radii}.

\paragraph{Non-thermal Component}
Assuming different spectral indices in modeling the none-thermal
spectral component also has a small effect on the inferred
radii \citep{heinke2014}.  The low counts in the spectra do not allow an
accurate measurement of this parameter; however, a range of values
have been explored in fitting the data.

\paragraph{Interstellar Extinction}
Because of the low temperature of the surface emission from qLMXBs,
the uncertainty in the interstellar extinction has a non-negligible
effect on the spectral analyses. Different amounts of interstellar
extinction have been assumed in different studies
\citep{guillot2013,lattimer2014}. A recent study explored different
models for the interstellar extinction in the analysis of the qLMXBs
in $\omega$~Cen and NGC~6397 and found statistically consistent
results with small differences in the central values but larger
differences in the uncertainties \citep{heinke2014}.  The relevant
uncertainties have been incorporated into the results shown in
Fig.~\ref{fig_spec_radii}.

\subsubsection{Thermonuclear Bursts}
Neutron stars in low-mass X-ray binaries show a phenomenon called
thermonuclear (Type-I) X-ray bursts, in which the accreted material
undergoes a helium flash that consumes the fuel that is spread over
the neutron star surface. The observed X-ray luminosity rises rapidly,
in a timescale of $\approx 1$~s, which corresponds to the diffusion
timescale from the burning layer to the neutron star surface. The
energy is then radiated away on a timescale of $15-50$~s for the
majority bursts during the so-called cooling tails. In a subset of the
cases, the luminosity reaches the Eddington luminosity where the
radiation forces match or exceed the gravitational forces and lift the
photosphere off of the neutron star surface. These photospheric radius
expansion (PRE) events have characteristic signatures where the
photosphere is observed to reach several times the neutron star radius
and the temperature has a second rise \citep{lewin1993}.

Several approaches have been developed and used to determine the
neutron star radius by making use of thermonuclear bursts.
\citet{van-paradijs1979} introduced using the apparent angular size
obtained during the cooling tails of the bursts as a way to obtain
correlated M-R constraints. Subsequently, \citet{fujimoto1986},
\citet{sztajno1987}, \citet{damen1990}, and \citet{ozel2006} discussed
different ways in which the degeneracy between the radius and the mass
can be broken through multiple spectroscopic measurements.

\citet{ozel2009}, \citet{guver2010a}, and a number of studies since
have used a combination of the apparent angular sizes, the Eddington
fluxes obtained from the PRE bursts, and the source distances to
measure the neutron star radius and mass. In particular, the apparent
angular sizes were measured from the flux and the temperature obtained
from time-resolved spectroscopy, while the Eddington limit is measured
at the moment when the photosphere touches back down onto the stellar
surface. The Eddington flux is related to stellar parameters $M$ and
$R$ and distance $D$ via
\begin{equation}
F_{\rm Edd}  = \frac{GMC}{k_{\rm es} D^2}\left(1-\frac{2GM}{Rc^2}
\right)^{1/2}, 
\label{eq_FEdd}
\end{equation}
where
\begin{equation}
k_{\rm es} \equiv 0.2 (1+X) \left[1+\left(\frac{kT_c}{38.8~{\rm keV}}\right)^{a_g} 
\left(1-\frac{2GM}{Rc^2}\right)^{-a_g/2}\right]^{-1}~{\rm cm}^2~{\rm g}^{-1}
\label{eq_es}
\end{equation}
is the electron scattering opacity and the last term in the equation
arises from temperature corrections to electron scattering
\citep{paczynski1983,suleimanov2012}. $X$ is the hydrogen mass
fraction of the atmosphere, 
\begin{equation}
a_g = 1.01 + 0.067 \left( \frac{g_{\rm eff}}{10^{14}~{\rm cm~s}^{-2}}\right),
\end{equation}
and the effective surface gravity is given by 
\begin{equation}
g_{\rm eff} = \frac{GM}{R^2} \left(1-\frac{2GM}{Rc^2} \right)^{-1/2}.
\end{equation}
Because the dependence of the Eddington flux on the stellar mass and
radius (Eq.~\ref{eq_FEdd}) is different from that of the apparent
angular size (Eq.~\ref{eq_rapp}), combining these two spectroscopic
measurements breaks the degeneracies between the stellar parameters
introduced by general relativistic effects.

Atmosphere models during bursts are used for the interpretation of
spectra and the correct measurement of effective temperatures. A
number of such models have been developed, addressing the
Comptonization of photons by the hot surface electrons with increasing
levels of sophistication, and have been calculated for a range of
surface compositions and effective gravities
\citep{london1986,madej2004,majczyna2005,suleimanov2012}.  The
deviations from a blackbody at the effective temperature of the
atmosphere are quantified by the so-called color correction factor
$f_c = T_{\rm c}/T_{\rm eff}$, which is then folded into these measurements.

Finally, the measured angular sizes are subject to additional general
relativistic corrections due to moderately high spin frequencies
observed in neutron stars with thermonuclear bursts. In particular,
the effects of the quadrupole moment and the ellipticity of the
neutron star on the neutron-star spacetime as well as the rotational
broadening of the thermal spectra can be calculated in the
Hartle-Thorne metric and lead to corrections in the inferred angular
size of the order of $\sim 10\%$ \citep{baubock2015}.

This approach has been applied to the neutron stars in the low-mass
X-ray binaries \fortyfive\ \citep{ozel2009},
\twenty\ \citep{guver2010a}, \oeight\ \citep{guver2010},
\thirtyone\ \citep{ozel2012}, \fortyeight\ \citep{guver2013}, and
\twentyfour\ \citep{ozel2015}.  \citet{ozel2015} performed an updated
analysis using new statistical methods, the theoretical corrections
discussed above, and uniformly incorporating systematic uncertainties
into the measurements. The resulting 68\% confidence contours over the
mass and radius for these six neutron stars are shown in
Fig~\ref{fig_spec_radii}. When combined, these measurements indicate
neutron star radii in the $9.8-11$~km range, which is consistent with
the results obtained from the analyses of qLMXBs.

Another approach that has been utilized for radius measurements is to
determine the spectral evolution of the color correction factor in PRE
bursts (see
\citealt{majczyna2005a,suleimanov2011,poutanen2014}). Because the
spectral distortions depend on the effective surface gravity $g_{\rm
  eff}$ and the emitted flux, the change in the color correction
factor with luminosity as the burst cools down from an Eddington flux
can provide a measure of the stellar mass and radius. The application
of this method to \twentyeight\ \citep{majczyna2005},
\twentyfour\ \citep{suleimanov2011} and \oeight\ \citep{poutanen2014}
led to radii measurements that range from being too small (in the case
of \twentyeight) to being too large (\oeight).  As we will discuss
later, these measurements are biased by the inherent limitations of
the data. The data selection used in the latter studies will also be
discussed below.

One final approach is to compare the observed properties of the
bursts, such as their recurrence and decay times, and their peak
fluxes, to the expectations from theoretical lightcurve models to
infer a surface redshift for the neutron star. These constraints can
then be combined with the spectral evolution during the burst cooling
tails to obtain constraints on the neutron star radius. Unfortunately,
the theoretical models in this case are not very predictive and fail
to explain the lightcurves of nearly any of the nearly 50 sources
extensively studied with RXTE \citep{heger2007,galloway2008}. Therefore, 
the application of this method has been limited to the one source, 
GS~1826$-$24, where models can approximately match the lightcurve 
properties, and resulted in the limits $R< 6.8-11.3$~km for 
$M<1.2-1.7 \; M_\odot$ \citep{zamfir2012}, also consistent with the 
results obtained from qLMXBs and with PRE bursts 
(see Fig.~\ref{fig_spec_radii}).

As with the qLMXBs in globular clusters, several sources of systematic
uncertainty exist in the spectroscopic measurements of thermonuclear
bursts. Their effects on the radius measurements have been studied in
recent years an will be discussed in some detail below. 

\paragraph{Distances}
Two of the three techniques discussed above rely on a measurement of
source distances to obtain neutron star radii. For the sources that
reside in globular clusters, the cluster distances are utilized.
These are subject to the same uncertainties as those used for the
qLMXB measurements. Other techniques have also been considered to
measure distances to bursting neutron stars
\citep{kuulkers2003,guver2010a,ozel2012}. In most cases, these
measurement techniques either yield reliable lower limits or source
distances with relatively large uncertainties, which dominate the
resulting uncertainties in the radius measurements.
 
\paragraph{Detection of Outliers}
As with all physical measurements, identification of outliers in
datasets that may contaminate the results is an important issue.
\citet{guver2012a} and \citet{guver2012b} developed a data-driven
Bayesian Gaussian-mixture approach to identify the outliers in the
spectroscopic burst data without any theoretical biases. The resulting
selected data were used in the radius measurements reported in
\citet{ozel2015}. In an alternative approach, \citet{poutanen2014} and
\citet{kajava2014} performed data selection by requiring that the
observations follow theoretical expectations. This approach resulted
in a selection of a different, and much smaller, subset of bursts and
led to the measurement of a stellar radius $R>12$~km for \oeight.  The
consistency of the burst evolution with theoretical expectations and
the limitations of the latter method when applied to RXTE data are
discussed in detail in \citet{ozel2015b}.

\subsection{Radii via Pulse Profile Modeling}

Neutron star radii can also be measured or constrained by analyzing
the properties of periodic brightness oscillations originating from
temperature anisotropies on the surface of a spinning neutron star.
The amplitudes and the spectra of the oscillation waveforms depend on
the neutron-star spacetime, which determines the strength of the
gravitational light bending the photons experience as they propagate
to an observer at infinity \citep{pechenick1983}, as well as on the
temperature profile on the stellar surface and on the beaming of the
emerging radiation. Using theoretical models for the emerging
radiation, the properties of the brightness oscillation can,
therefore, be used to probe the stellar spacetime and to measure its
radius and mass.

The theoretical work on neutron-star spacetimes has enabled
increasingly more precise calculations of the effects of the
gravitational lensing on the surface photons, which are used for the
analyses of the waveforms. Earlier approximations appropriate for
non-spinning neutron stars \citep{pechenick1983} were supplemented by
adding in the effects of Doppler shifts and aberration
\citep{miller1998a,muno2002,poutanen2006} or those of frame dragging
\citep{braje2000} for slow spins up to 300~Hz. At even faster spins,
the oblateness of the neutron star and the quadrupole moment of its
spacetime on the photon trajectories have been accurately accounted
for \citep{morsink2007,psaltis2014a}. For stars near break-up
($>1$~kHz), fully numerical solutions of the spacetimes become
necessary \citep{cook1994,stergioulas1995,cadeau2007}.

The waveforms depend also on the location and the size of the
hotspots, the beaming pattern of the radiation emitted from the
stellar surface, and the line of sight of the observer with respect to
the rotation axis.  Modeling the shapes and amplitudes of the
waveforms, therefore, can yield constraints on a combination of all of
these parameters and result in measurements of these parameters with
correlated uncertainties. The challenge here is to reduce the number
of unknowns and to eliminate the correlated uncertainties between the
parameters in order to obtain a measurement of the neutron star
radius. The size of the hotspot has a minor effect when it is
sufficiently small \citep{baubock2015}), so this parameter can be
eliminated in some cases. The beaming of radiation can also be
calculated from atmosphere models.

\begin{figure}
\includegraphics[height=3in]{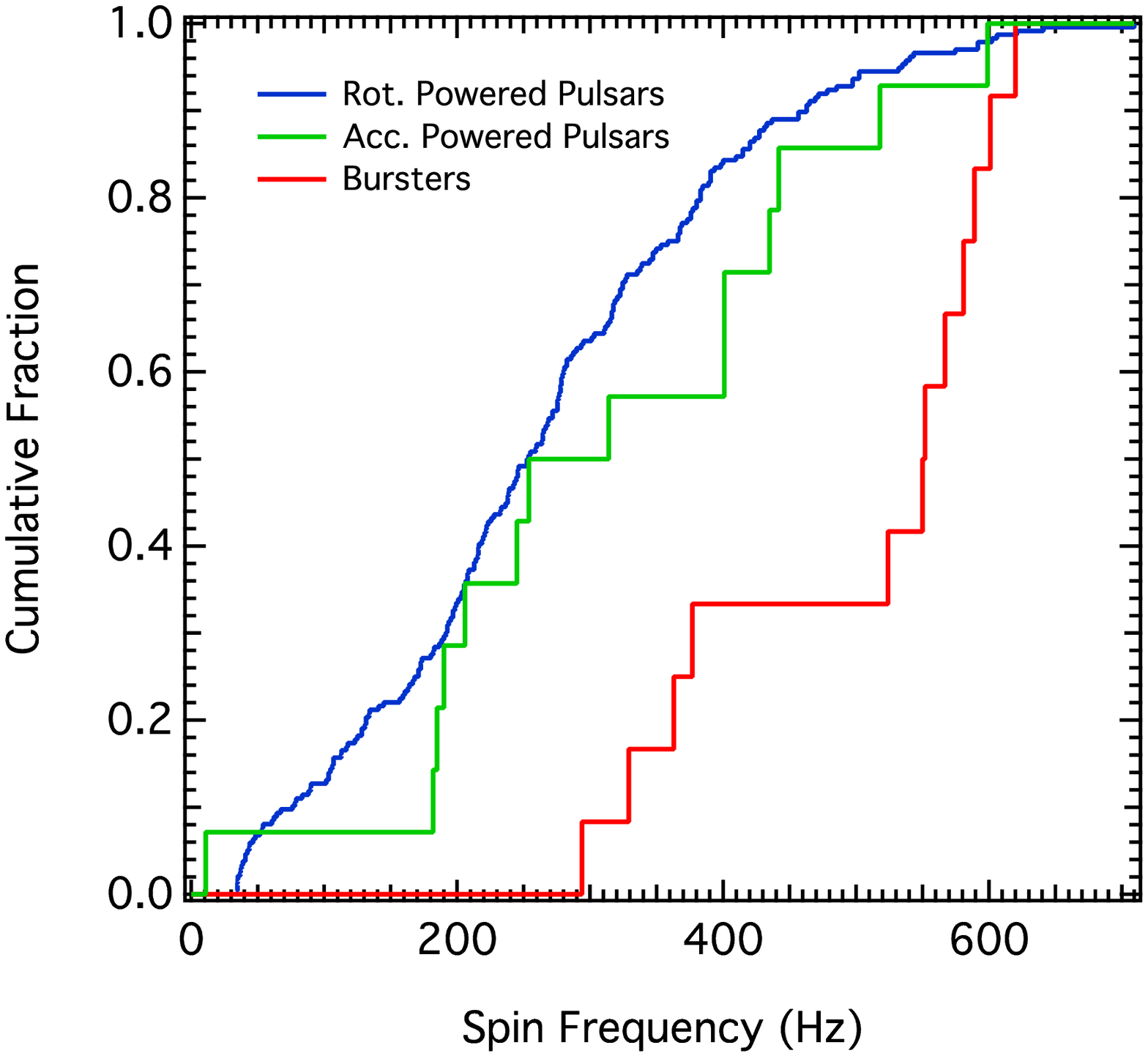}
\caption{The cumulative distribution of spin frequencies of
  rotation-powered millisecond pulsars, accretion powered millisecond
  pulsars, and accreting neutron stars that show thermonuclear burst
  oscillations.}
\label{fig_spins}
\end{figure}

Analysis of the oscillation waveforms have been performed to explore
the properties of numerous types of neutron stars. While this method
yielded useful results on the surface emission properties of slow
pulsars \citep{page1995} and magnetars \citep{dedeo2001,ozel2001},
constraints on the neutron-star radius and spacetime come, in
particular, from the analysis of oscillations from accretion-powered
millisecond pulsars, millisecond rotation-powered pulsars, and
thermonuclear X-ray bursters. These last three classes have
distributions of spin frequencies that differ from one another, as
shown in Fig.~\ref{fig_spins}. Therefore, spin effects need to be
incorporated to different levels when modeling each population. We
discuss the radius measurements from each of these classes
individually below.

\subsubsection{Rotation Powered Pulsars}

Even though the emission from rotation-powered millisecond pulsars
(MSPs) is largely non-thermal and is dominated by their
magnetospheres, a number of sources show a clearly detected thermal
component in the soft X-rays (e.g.,
\citealt{grindlay2002,zavlin2006,zavlin2007,bogdanov2006}). X-ray data
from a number of MSPs, obtained with ROSAT, Chandra, and XMM-Newton,
have been analyzed using hydrogen atmosphere models for the thermal
emission from a polar cap (see, e.g., Zavlin \& Pavlov 1998). In these
models, pulse profiles are calculated based on the beaming of
radiation predicted by the theoretical models (see \citealt{ozel2013}
for details).

\begin{figure}
\includegraphics[height=3in]{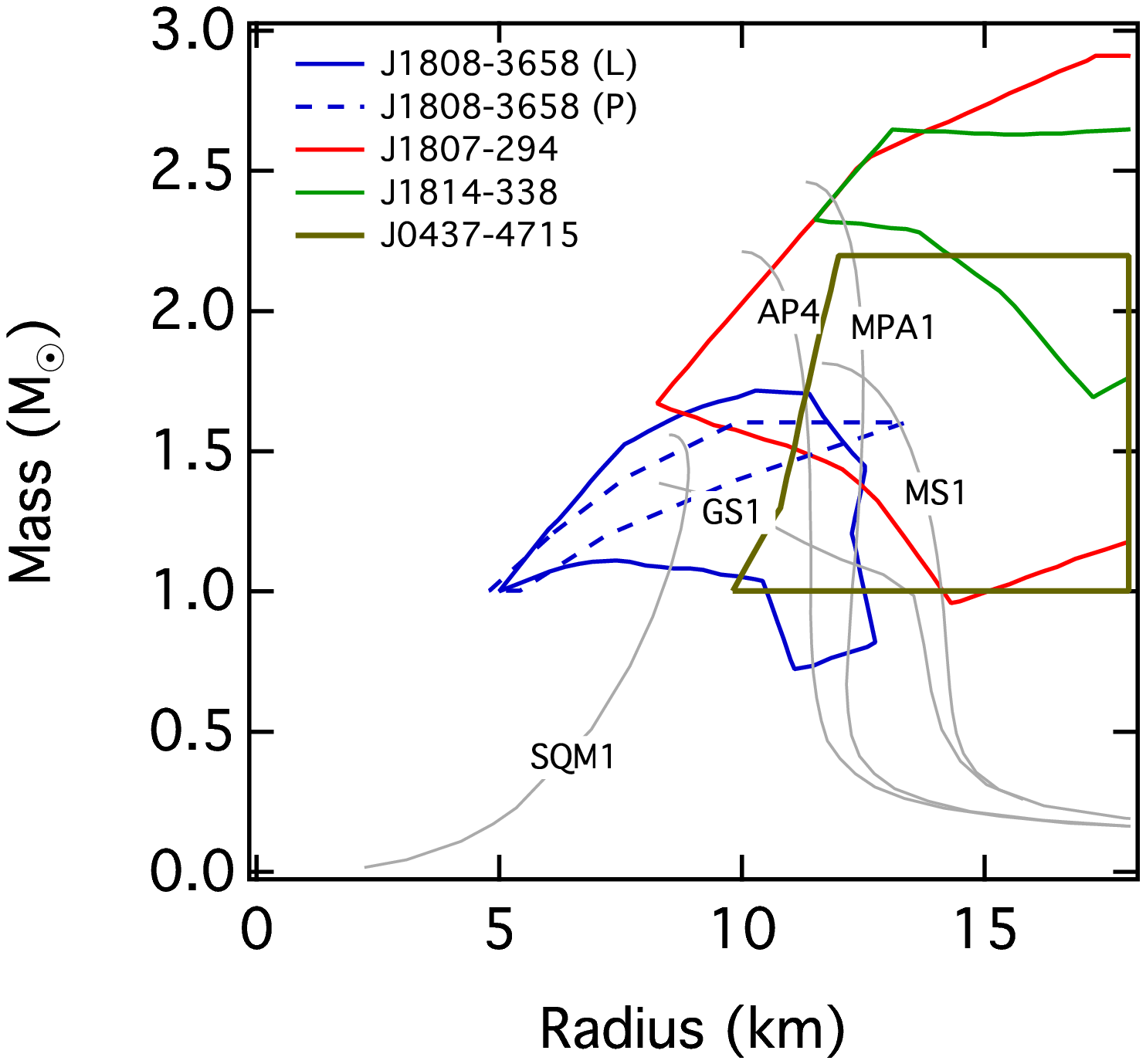}
\caption{The radii constraints obtained from analysis of the waveforms
  from accretion-powered \citep{poutanen2003,leahy2008,leahy2009,leahy2011} 
  and rotation-powered millisecond pulsars \citep{bogdanov2013}.  Two
  different analyses of the SAX J1808.4-3658 data by Poutanen et
  al. (2003, denoted by P) and Leahy et al.\ (2008, denoted by L) are
  included.}
\label{fig_pulse_radii}
\end{figure}

An analysis of PSR J0437$-$4715, assuming a mass of 1.4~$M_\odot$
yielded bounds on the radius in the $R=6.8-13.8$~km range (90\% C.L.;
\citealt{bogdanov2007}). A subsequent measurement of the pulsar mass
at 1.76~$M_\odot$ \citep{verbiest2008} increased the lower limit on
the neutron star radius to $R>11.1$~km (99.9\% C.L.; see
Fig.~\ref{fig_pulse_radii} and \citealt{bogdanov2013}), but this
result is likely to be revised again in view of the most recent
measurement of the pulsar mass (see Table~\ref{tab2}). Similar
analyses for pulsars J0030+0451 and J2124$-$3358 lead to lower limits
on their radii of 10.7~km (95\% C.L.) and 7.8~km (68\% C.L.),
respectively, assuming a pulsar mass of 1.4~$M_\odot$.
 
There are a number of complexities that affect the interpretation of
the pulsar waveform data from MSPs. First, the non-thermal
magnetospheric emission is subtracted from the total observed emission
in the X-rays to allow for a modeling of the thermal surface component
from the polar caps. The thermal component itself is typically modeled
with two separate regions with different temperatures, each emitting
with a hydrogen atmosphere spectrum \citep{zavlin2006,bogdanov2007}.
The pulse profiles have been modeled with two polar caps but the
waveforms have required a non-antipodal geometry and led to additional
uncertainties from introducing an additional offset parameter. The
footprints of polar caps in the studies performed to date have assumed
circular caps but the effect of the shape of the polar cap on the
waveforms have not yet been explored (see, however,
\citealt{baubock2015} for the effect of the unknown shape and size of
the polar caps). Finally, the angles that specify the observer's
line-of-sight and the colatitude of the polar caps with respect to the
stellar spin axis have been assumed to be known independently. The
uncertainties introduced by the errors in these angles have not been
factored into the quoted radius constraints. Future work is needed to
assess the effects of these assumptions and to quantify the additional
resulting uncertainties in the radius measurements.

\subsubsection{Accretion Powered Pulsars} 

Thermal emission is observed in a number of accretion-powered
millisecond X-ray pulsars (AMSPs) in the soft X-rays. It originates
from the base of the accretion column, which is then Comptonized in
that column (see section 3.3 of Poutanen \& Gierlinski 2003).

A number of studies have modeled the emission from the X-ray emitting
regions at the bases of the accretion columns
\citep{poutanen2003,lamb2009}. The thermal photons emitted from the
surface are Compton scattered in the accretion column, affecting the
observed pulse amplitudes. \citet{lamb2009} also explored the effect
of the location of the X-ray emitting regions and considered models
where the base of the accretion column wobble around the spin axes.
These models help interpret the AMSP timing data but also explore
uncertainties in the mass-radius constraints obtained from fitting the
waveforms.

This method has been applied to several accreting millisecond X-ray
pulsars: SAX J1808.4$-$3658 \citep{poutanen2003,leahy2008,morsink2011}, 
XTE J1807$-$294 \citep{leahy2011}, and XTE J1814$-$338 \citep{leahy2009}. 
Fig.~\ref{fig_pulse_radii} shows the inferred constraints on the mass
and radius of the neutron stars in these studies, where the contours
depict 99\% confidence levels. The large uncertainties in the
mass-radius measurements shown in this figure reflect the influence of
the various geometric factors discussed above, which are difficult to
constrain.

\subsubsection{Thermonuclear Burst Oscillations}

Periodic flux oscillations with large amplitudes originating from the
surfaces of neutron stars have also been detected during thermonuclear
X-ray bursts \citep{strohmayer1996}. They originate from temperature
non-uniformities on the stellar surface that occur during the nuclear
burning (see \citealt{watts2012} for a review). Modeling the pulse
profiles of these burst oscillations are also good probes of the
stellar gravity and leads to constraints on the neutron star
compactness \citep{strohmayer1997,miller1998}. In particular, the
amplitude of the oscillations, the deviations from a sinusoidal
waveform, and their dependence on photon energy can be probes of the
neutron star radius and mass \citep{weinberg2001,muno2002,muno2003}.

This technique was applied to oscillations observed from 4U~1636$-$536
\citep{nath2002} but led to no significant constraints on the neutron
star radius or compactness. For the neutron star in XTE J1814$-$338,
the analysis of the waveforms of burst oscillations resulted in the
limit on the neutron star compactness $R/M > 4.2$ \citep{bhattacharyya2005}.  
Future observations can make use of more properties of the observed waveforms, 
such as their energy dependence and the amplitudes of the higher harmonics 
\citep{psaltis2014}, and the prospects are more comprehensively discussed 
in Watts et al.\ (2015).

\section{The Neutron Star Equation of State}

In addition to characterizing neutron star populations as well as
understanding their formation and evolution, one of the ultimate goals
of mass-radius measurements is, of course, to constrain the ultradense
matter equation of state (EoS). The microphysics, as described by the
EoS, is connected to the macroscopic properties of the neutron star
via the Tolman-Oppenheimer-Volkoff equations, which provide the direct
relation that is necessary to use astrophysical observations to
constrain nuclear physics at very high densities. In this section, we
will describe this relation, a variety of theoretical approaches to as
well as the low-energy experimental constraints on the EoS, and the
current measurement of the EoS beyond nuclear saturation density from
neutron star observations.

\subsection{Neutron Star Structure}

The structure of relativistic stars and their mass-radius relation are
determined by integrating the relativistic structure equations, also
known as the Tolman-Oppenheimer-Volkoff (TOV) equations
\citep{oppenheimer1939,tolman1934,tolman1939}
\begin{eqnarray}
\frac{dp}{dr} &=& - \frac{G}{c^2} \frac{(p+\epsilon)(m+4\pi r^3 p/c^2)}{r^2[1-2Gm/(rc^2)]} \nonumber \\
\frac{dm}{dr} &=& 4\pi r^2 \frac{\epsilon}{c^2}, 
\end{eqnarray}
where $p$ is the pressure, $\epsilon$ is the energy density, $r$ is
the radius, and $m$ is the enclosed mass, from the center $r=0$ where
$p=p_c$ and $\epsilon=\epsilon_c$ to the surface at $r=R$, where $p=0$
and the enclosed mass is equal to the total mass $m=M$. (At moderate
spin frequencies $f_{\rm NS} \lesssim 800$~Hz, rotational corrections
appear at $<$ few \% level;
see \citealt{cook1994,stergioulas1995}). To complete this set of
equations, one needs to specify the pressure-energy density relation,
referred to as the equation of state, appropriate for neutron-star
matter.

Before we proceed with a discussion of the cold, dense matter EoS, a
few remarks on the relevant quantities and units are in order. While
the energy density $\epsilon$ enters the TOV equations, most nuclear
EoS calculations also use or refer to the baryon number density $n$
(in units of ${\rm fm}^{-3} \equiv 10^{39}{\rm cm}^{-3}$) or to the
mass density $\rho$ (in units of g~cm$^{-3}$). The nuclear saturation
density can then be equivalently expressed as a mass density $\simeq
2.8 \times 10^{14}$~g~cm$^{-3}$ or a particle density $\simeq
0.16~{\rm fm}^{-3}$.  Similarly, the pressure is typically expressed
in units of MeV~fm$^{-3} \equiv 1.6 \times 10^{33}$~dyne~cm$^{-2}$.
 
In neutron star cores, matter is cold and ultradense, characterized by
large chemical potentials and a significant asymmetry in the numbers
of neutrons and protons. Indeed, the condition of $\beta$ equilibrium, 
given by
\begin{equation}
\left( \frac{\partial \epsilon /n}{\partial Y_e} \right)_n = 
\mu_e + \mu_p - \mu_n = 0
\end{equation}
points to proton fractions of $x \lesssim 10\%$, unlike normal nuclei
which possess nearly equal numbers of protons and neutrons (referred
to as symmetric matter). In the equation above, $\epsilon/n$ is the
energy per baryon, $\mu_e, \mu_p$, and $\mu_n$ are the chemical
potentials of the electrons, protons, and neutrons, respectively, and
$Y_e$ is the number of electrons per baryon. The densities in the
cores can reach $\sim 10 \rho_{\rm sat}$. Given that neutrons
geometrically overlap at $\sim 4 \rho_{\rm sat}$, at these very high
densities, matter may not be composed of nucleons alone but may
contain a rich variety of hadronic degrees of freedom. With increasing
overlap between nucleons, transitions to non-nucleonic states of
matter are expected. For example, a gradual onset of quark degrees of
freedom and quark constituents propagating throughout the system
become likely (e.g., \citealt{alford2005a,kojo2015}). In addition,
Bose-Einstein condensates of mesons, such as pion
\citep{akmal1998,mukherjee2009,pandharipande1995} or kaon condensates
\citep{kaplan1986,brown1992,glendenning1999} can enter at higher
densities. It is also possible for ultradense matter to contain
strange quarks in forms other than kaons, as weak interactions in the
cold neutron-star cores can lead to states of matter with a high
degree of strangeness. Some of the possibilities considered to date
include the presence of hyperons
\citep{ambartsumyan1960,glendenning1991,schulze1995}, hybrid stars
containing free quarks \citep{collins1975}, or color superconducting
phases \citep{alford2003,alford2005}. Moreover, self-bound stars
consisting of strange quark matter have also been proposed
\citep{farhi1984,haensel1986,alcock1986}. These models are based on
the assumption that strange quark matter is the ultimate ground state
of matter \citep{witten1984} and they predict stellar radii that grow
with the stellar mass.

First principles QCD calculations are not yet able to describe the
interactions of such a system at several times the nuclear saturation
density (but there is significant recent progress in describing the
quark matter EoS at even higher densities and extrapolating down to
lower densities; see, e.g., the discussion and results in
\citealt{kojo2015} and \citealt{kurkela2014}). Because of that, a
number of different approaches have been followed in the calculations
of the equation of state of neutron star matter. A comprehensive
overview of the methods and the details of the nuclear physics can be
found in \citet{glendenning2000}, \citet{lattimer2001}, and
\citet{baldo2012}. Here, we provide a summary of the basic methods as
well as a compilation of some representative model equations of state.

One common effective field theory motivated approach relies on
expanding the nuclear Hamiltonians into two-, three-,...  n-body
static potential terms, assuming that the forces between particles can
be described via static few-body potentials. The two-body potential is
determined in the vicinity of $\rho_{\rm sat}$ for symmetric matter
using nucleon-nucleon scattering data below 350~MeV and the properties
of light nuclei, and the contributions from the three-body potentials
are added \citep{akmal1998, morales2002,hebeler2010,gandolfi2012}. The
resulting equations of state are particularly sensitive to the
three-nucleon interactions because of the near-cancellation between
the kinetic term and the two-body term in these potentials. In
addition, such expansions are not valid at $\rho \gg \rho_{\rm sat}$
because the parameter measuring the relative importance of three and
higher body forces is $\sim \rho/(2m_\pi)^3 \sim 0.35\rho/\rho_{\rm
  sat}$.

In these approaches, it is also common to make use of nuclear
Hamiltonians that are experimentally constrained for symmetric matter
around nuclear density and express the energy per baryon at higher
density and at higher neutron fraction as a parametric expansion in
terms of the ratio $u \equiv \rho/\rho_{\rm sat}$ and the neutron
excess $1-2x$:
\begin{equation}
\frac{\epsilon (u,x)}{n}=-B+{K_o\over18}(u-1)^2+{K_o^\prime\over162}(u-1)^3 +S_2(u)(1-2x)^2 
\label{eq:expansion}
\end{equation}
In this expansion, $B\simeq 16$~MeV is the bulk binding energy of
symmetric matter at nuclear saturation density and $K_o$ and
$K_o^\prime$ are the nuclear incompressibility and skewness parameters
and we only show terms up to quadratic order. It is also customary to
call the quadratic term in neutron excess the ``symmetry energy'' and
express it in terms of two parameters $S_v$, the symmetry energy at
nuclear density
\begin{equation}
S_v = S_2(1),
\label{eq:Sparam}
\end{equation}
and $L$, which is related to the derivative of the symmetry energy
with respect to density
\begin{equation}
L = 3u \left(\frac{dS_2}{du}\right)_{u=1}. 
\label{eq:Lparam}
\end{equation}
Naturally, at high density and high neutron fraction, these approaches
also enter a regime where the expansions are not valid.

Alternatives to this approach include one that is based on
field-theoretical calculations of constituents interacting via meson
exchange \citep{muller1996}.  A third approach involves microscopic ab
initio calculations based on the Brueckner- Hartree-Fock model
\citep{baldo1997} or its relativistic counterpart, the
Dirac-Brueckner-Hartree-Fock model \citep{muther1987}, or its
modifications using unconventional Skyrme forces
(\citealt{goriely2010}; see also \citealt{potekhin2013} for analytical
representations based on this equation of state).

\begin{figure}
\begin{minipage}{2.8in}
\includegraphics[height=2.7in, width=2.7in]{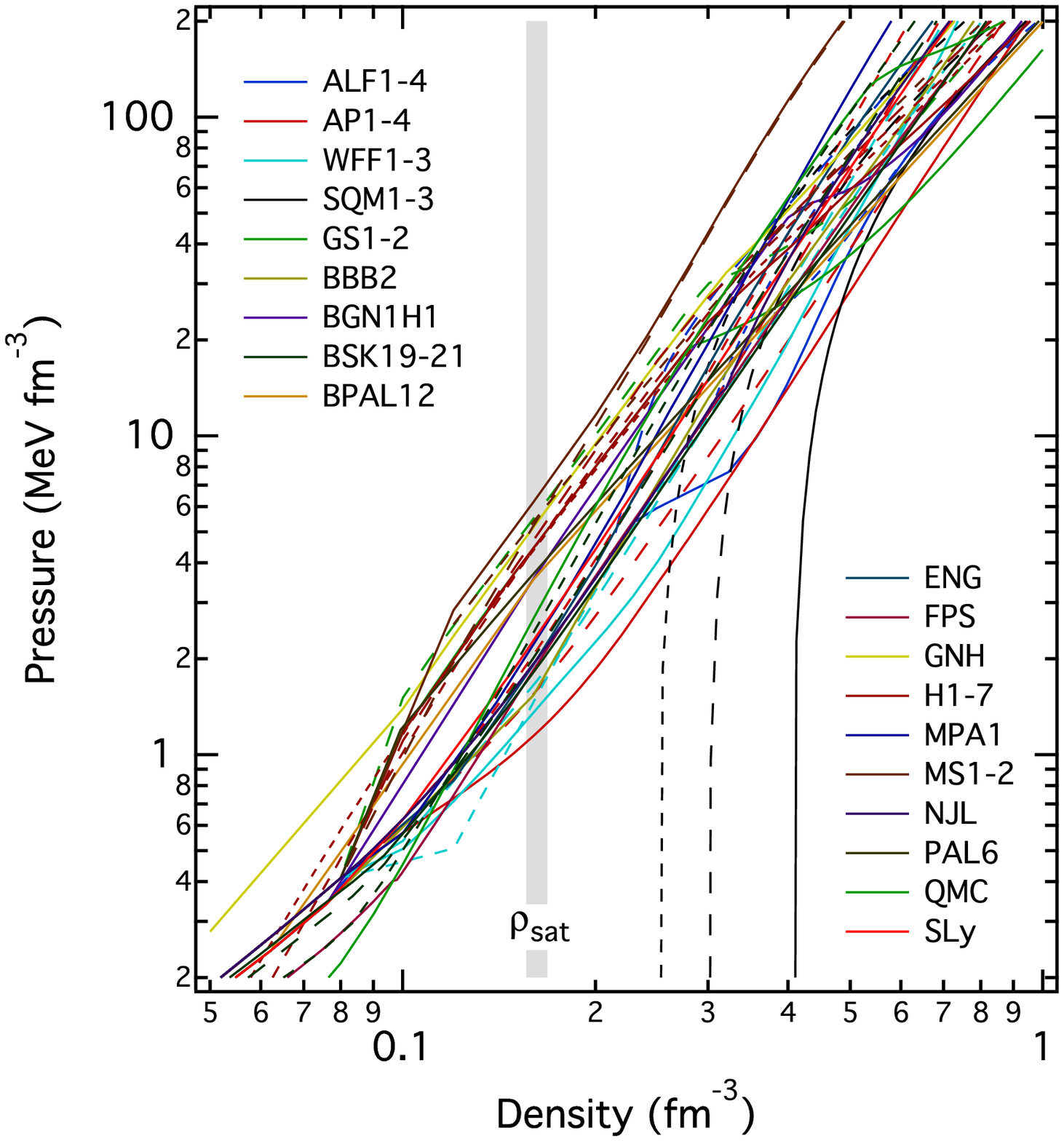}
\end{minipage}
\begin{minipage}{3.0in}
\includegraphics[height=2.7in, width=2.7in]{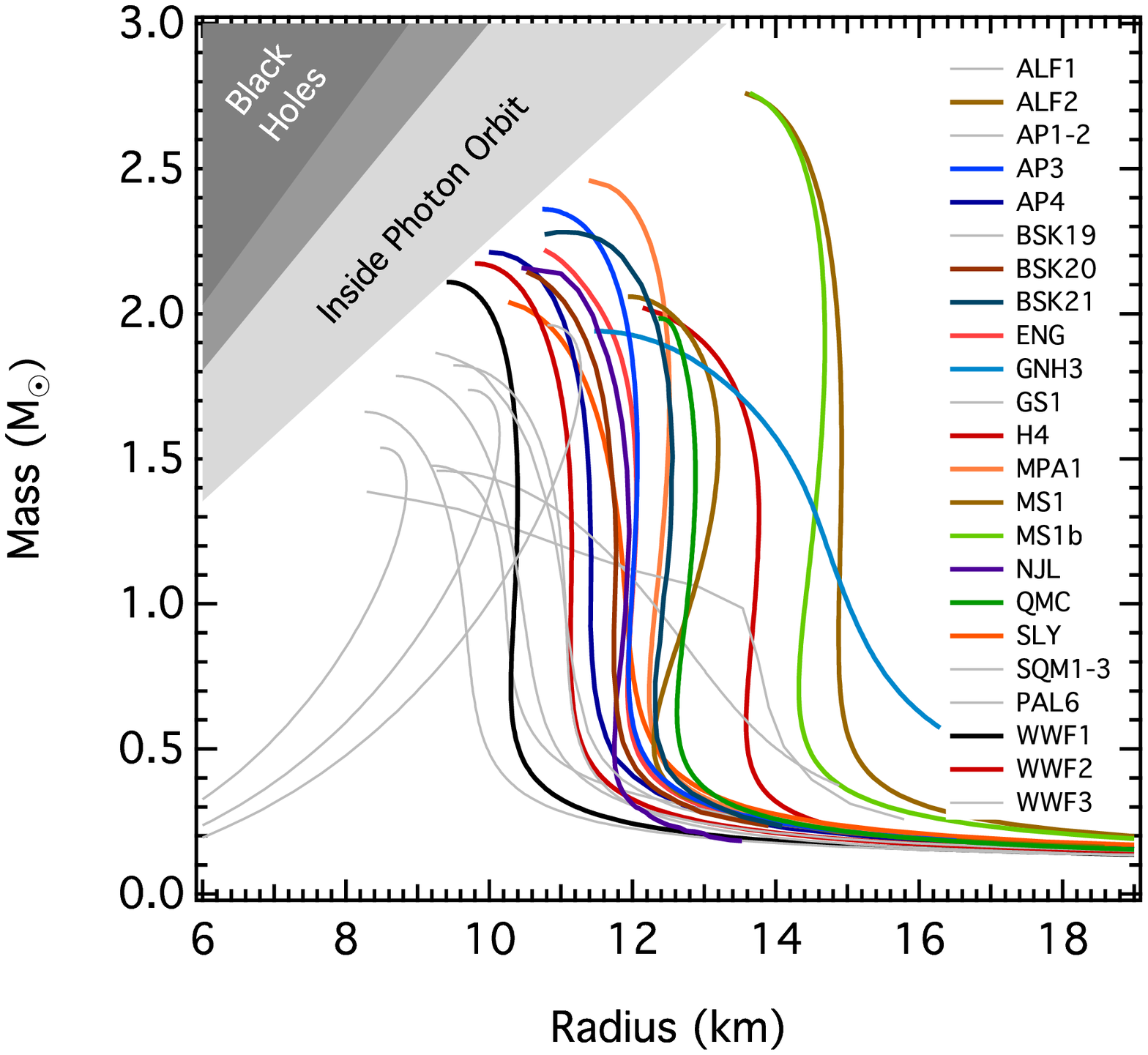}
\end{minipage}
\caption{(Left) A large sample of proposed equations of state
  calculated under different physical assumptions and using a range of
  computational approaches. See the text for the descriptions of the
  equations of state, the acronyms, and the references. (Right) The
  mass-radius curves corresponding to the equations of state shown in
  the left panel.}
\label{fig_EoS}
\end{figure}

Fig.\ref{fig_EoS} (left panel) shows a large sample of currently
proposed equations of state, spanning a density range between $\sim
0.1 - 8 \rho_{\rm sat}$. The wide range of pressures $P$ predicted
under different assumptions are indicative of the wide open questions
in physics in this regime. The acronyms and the references for the
various equations of state can be found in \citet{read2009}. We also
include in this figure {\it (i)}\ the recent equation of state labeled
NJL \citep{kojo2015} based on a smooth interpolation in pressure
vs. baryon chemical potential of a nucleonic equation of state (APR)
at densities below $\sim \rho_{\rm sat}$ with a quark matter equation
of state at densities above $\sim 5-7 \rho_{\rm sat}$, (ii) the BSK
equation of state of \citet{goriely2010} and \citet{potekhin2013}, and
(iii) the QMC equation of state of \citet{stone2007}.

The mass-radius relations corresponding to the equations of state in
Fig.~\ref{fig_EoS} are shown in the right panel. The curves that do
not reach the observed $\sim 2 \; M_\odot$ mass limit are denoted in
light grey in this figure. Also note that the astrophysically relevant
parts of these curves lie above $\sim 1 \; M_\odot$. An important
characteristic of many of these curves is that the radius remains
nearly constant for the astrophysically relevant range of masses. The
notable exceptions are the self-bound strange stars (e.g., SQM), where
the radius increases with increasing mass, and stars with condensates
(e.g., GS1-2, GM, PS) where the radius decreases with mass past the
point where the central density reaches the critical one where the
phase transition occurs. The mass-radius curves are also characterized
by a maximum mass beyond which there are no stable solutions. In
general, equations of state with relatively higher pressures at
densities above $\sim 4 \; \rho_{\rm sat}$ have higher maximum masses.
The presence of non-nucleonic phases, such as hyperons or condensates,
reduces the pressure (referred to as softening the equation of state)
lead to smaller maximum masses.

In the remainder of this section, we will discuss how neutron star
masses and radii can be used to pin down the ultradense matter
equation of state, the methodologies developed towards this goal, and
the current state of the measurements. However, we first briefly
describe the constraints on the nuclear EoS at nuclear density from
low energy experiments.

\subsection{Constraints on the EoS from Low Energy Experiments}
    
\begin{figure}
\includegraphics[height=3in]{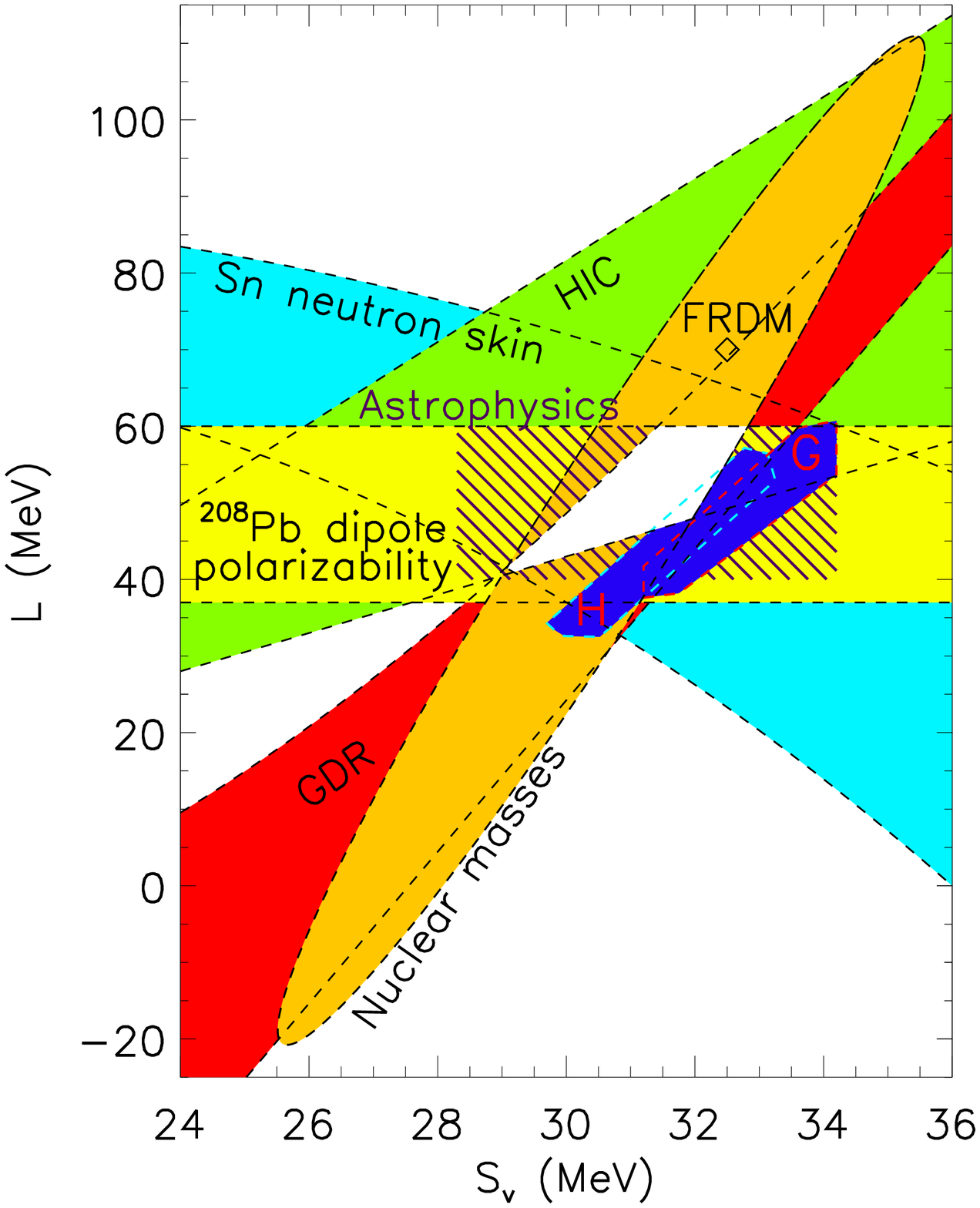}
\caption{The experimental constraints on the symmetry energy
  parameters $L$ and $S_v$ from experiments performed near nuclear
  density. These include the neutron skin thickness of neutron-rich Sn
  isotopes (shown in blue), giant dipole resonances (red), the dipole
  polarizability of $^{208}Pb$ (light yellow), heavy-ion collisions
  (green), as well as nuclear masses (dark yellow). The theoretical
  inferences from Gandolfi (2012, G) and Hebeler (2010, H), as well as
  an earlier astrophysical inference from \citealt{lattimer2012} are
  also indicated on this figure as dark blue and hatched regions,
  respectively. Reproduced with permission from
  \citealt{lattimer2012}. }
\label{fig:exp_constr}
\end{figure}

For symmetric matter (i.e., nuclei containing roughly equal number of
neutrons and protons) near the nuclear saturation density, there is a
range of experimental constraints. Most robustly, two-body potentials
can be inferred from nucleon-nucleon scattering data below 350 MeV and
the properties of light nuclei \citep{akmal1998,morales2002}.

The other significant constraints that arise from these experiments
and are relevant for the neutron-star equation of state are often
expressed in terms of the symmetry energy parameters: $S_v$ and $L$
(see eq'ns \ref{eq:Sparam} and \ref{eq:Lparam} in the previous section
as well as the discussion in \citealt{lattimer2012}). The experiments
that yield the most accurate data and the least model-dependent
results involve fitting nuclear masses and charge radii
\citep{klupfel2009,kortelainen2010}. Nevertheless, the symmetry
parameters that can be extracted from such data are highly correlated,
as shown in Fig.~\ref{fig:exp_constr}.

Neutron-rich matter can also be probed by measuring the neutron skin
thickness of heavy nuclei. Studies within both the mean-field theory
and the liquid droplet model frameworks have shown that the neutron
skin thickness, defined as the difference of their neutron and proton
root-mean-squared radii
\begin{equation}
\Delta r_{np} = <r_n^2>^{1/2} - <r_p^2>^{1/2}
\end{equation}
is a sensitive function of $S_v$ and $L$ and, thus, serves as a good
probe of the symmetry energy
\citep{centelles2009,chen2010,roca-maza2011}.  The neutron skin
thickness was measured by a variety of experiments for $\approx 20$
neutron-rich Sn isotopes with $\sim 30-50\%$
uncertainties. \citet{chen2010} used these measurements to place
additional constraints on the symmetry energy parameters (see
Fig.~\ref{fig:exp_constr}). In addition, the neutron skin thickness
has been determined by the PREX experiment for $^{208}Pb$ using a
technique called parity-violating electron scattering
\citep{abrahamyan2012}: because the weak-vector meson that mediates
this interaction couples preferentially to neutrons, the approach
allows a measurement of the neutron form factor, which can then be
extrapolated to neutron skin thickness.

Another observable that places significant constraints on $L$ come
from the measurement of the dipole polarizability of $^{208}Pb$ and the
giant dipole resonances. The giant dipole resonance, measured from the
response of the neutron-rich nuclei to an externally applied electric
field, serves as a measure of the symmetry energy
\citep{trippa2008,tamii2011,piekarewicz2012}.  The resulting
constraints on $S_v$ and $L$ are also shown in
Fig.~\ref{fig:exp_constr}.

The most indirect experimental probe of the EoS of cold, neutron-rich,
dense matter emerge from low-energy heavy-ion collisions. There are
large uncertainties in extrapolating from the properties of hot dense
matter generated in these collisions to zero temperature and to
extreme neutron-richness \citep{tsang2009}.  For completeness,
Fig.~\ref{fig:exp_constr} also includes these constraints.

While none of these experiments are as powerful and direct probes of
the neutron star EoS as the measurement of the macroscopic properties
of neutron stars (as will be shown in the next section), they serve as
complementary tests of the properties of dense matter.  At a minimum,
the pressure obtained from the two-nucleon interaction serves as a
lower limit to the pressure of neutron-star matter at $\sim 1-2 \;
\rho_{\rm sat}$. This is because three-body interactions in pure
neutron matter are always repulsive. Considering the expansion of the
interaction in terms of two- and three-body potentials and using the
contribution of the two-body potential (Argonne AV8 or AV18) yields a
lower bound of $P = 7.56$~MeV~fm$^{-3}$ at $\rho=1.85 \; \rho_{\rm
  sat}$ \citep{ozel2015}. This constraint has been used in conjuction
with the neutron star radius measurements to determine the ultradense
matter EoS, as we will discuss in Section~4.4.

\subsection{From Radius and Mass Measurements to the Neutron Star EoS}

Given the Tolman-Oppenheimer-Volkoff equations, the EoS of neutron
star matter determines the macroscopic properties of the stars and, in
particular, their masses and radii. In fact, the solutions provide a
unique map between the microscopic pressure-density relation
($P(\rho)$ or $P(\epsilon)$) and the macroscopic mass-radius relation
of stars \citep{lindblom1992}.


This unique mapping can, in principle be exploited to infer the EoS
from astrophysical measurements of their masses and radii.  However, a
formal inversion of this mapping requires a masurement of the entire
mass-radius curve, i.e., the measurement of radii for neutron stars
that span the range of stellar masses between $\sim 0.2-2
M_\odot$. This requirement severely limits the applicability of this
direct inversion because there is no astrophysical formation channel
for neutron stars with masses that are much smaller than the
Chandrasekhar mass of the progenitor cores and neither are such stars
observed (see Section~2).

Despite this difficulty, the connection between the mass-radius
relations and the EoS turn out to have certain characteristics that
make it possible to infer them from a more sparse sampling of neutron
star radii. In particular, the maximum mass is determined primarily by
the behavior of the equation of state at the highest densities ($\sim
7-8 \; \rho_{\rm sat}$; \citealt{ozel2009}), while the radius depends
primarily on the pressure at $\approx 2 \; \rho_{\rm sat}$
\citep{lattimer2001}. The slope of the mass-radius relation (i.e.,
whether R increases or decreases with M), in the mass range of
astrophysical interest, depends largely on the pressure at $\sim 4 \;
\rho_{\rm sat}$ \citep{ozel2009}. The fact that only a small number of
densities determines the macroscopic properties of neutron stars means
that their various higher order moments cannot be completely
independent from each other. Indeed, a number of universal relations
have been found between various higher-order moments of neutron stars
\citep{yagi2013,baubock2013a,yagi2014}, the origins of which have not
yet been fully understood.

These properties make it possible to infer the pressure of ultradense
matter at a few appropriately chosen densities above $\rho_{\rm sat}$
instead of mapping out the full functional form of the EoS from
astrophysical observations. The optimal densities for sampling and the
parametric forms that lead to the smallest uncertainties in the EoS
determination were explored in detail by \citet{read2009} and
\citet{ozel2009}.  These studies showed that the pressures measured at
1.85 ($P_1$), 3.7 ($P_2$), and 7.4~$\rho_{\rm sat}$ ($P_3$) and
connected by piecewise polynomials can represent more than forty
proposed EoS included in those studies to within few percent
uncertainty. These led to the realization that the measurement of the
masses and radii of even a small number of neutron stars can provide
significant input to the microphysics calculations, which was used in
the subsequent studies \citep{ozel2010a,steiner2010,guillot2014}.

The mass of the heaviest neutron star measured to date can provide
additional direct constraints on the EoS. At a minimum, the EoS that
have maximum masses that fall below the the most massive neutron star
can be ruled out
\citep{demorest2010,ozel2010,antoniadis2013}. However, more stringent
constraints are achieved by combining the floor on the maximum mass
with radius measurements \citep{steiner2010,ozel2015}. Because the EoS
that produce smaller radii generally tend to have smaller maximum
masses, combining these measurements yield smaller allowed confidence
regions in the EoS parameters.

Bayesian techniques have been developed
\citep{steiner2010,steiner2013,ozel2015} to combine radius
measurements, the maximum mass constraint, as well as the data
obtained from the low-energy experiments discussed in the previous
section. These techniques yield likelihoods over the EoS parameters
$P_1, P_2$ and $P_3$ from the mass-radius likelihoods $P(M_i, R_i)$
for $i$ sources, while allowing a variety of priors and other
measurements to be incorporated (see \citealt{ozel2015} and references
therein for details).

The priors over $P_1, P_2$, and $P_3$ are then specified using
numerous physical and observational constraints. First, for the EoS to
be microscopically stable, it is necessary to require $P_3 \ge P_2 \ge
P_1$, and that $P_1$ be greater than or equal to the pressure of
matter at $\rho_0=10^{14}$~g~cm$^{-3}$ (see \citealt{ozel2009} for the
particular specification using the SLy EoS). Second, for
the EoS to be causal, the condition
\begin{equation}
c_s^2=\frac{\partial P}{\partial \epsilon} \le c^2,
\end{equation}
where, $c_s$ is the sound speed and $\epsilon$ is the energy density,
can be imposed at all densities. Third, to incorporate the mass
requirement, the maximum stable mass for each EoS corresponding to a
$P_1, P_2, P_3$ triplet needs to exceed $1.97 M_\odot$, consistent
within $1\sigma$ with the heaviest neutron stars observed to date.

The results from the laboratory experiments can be folded in as a
requirement on the pressure at the lowest density $P_1$ or as
constraints on the symmetry energy parameters $L$ and $S_v$, which can
be translated into constraints on $P_i$. As discussed in Section 4.2,
the least model-dependent one of those requirements is a lower limit
on $P_1 = 7.56$~MeV~fm$^{-3}$ such that the EoS is consistent with the
most robust laboratory experiments and low density calculations, i.e.,
with the APR EoS for pure neutron matter \citep{akmal1998}.  Finally,
priors that are flat in $\log P_1$, $\log P_2$, and $\log P_3$ or that
are flat in $P_1$, $P_2$, and $P_3$ can be chosen.  The sensitivity of
the Bayesian inference of the EoS parameters on these conditions and
priors is discussed in \citet{steiner2013} and \citet{ozel2015}.

\subsection{Current constraints on the cold neutron matter EoS}

In recent studies, the radius results summarized in Section 3 have
been used in conjunction with the Bayesian parameter inference
techniques to place strong constraints on the neutron star EoS. The
radius data include the fourteen sources shown in Figure~4.  (Note
that these omit the analysis of \twentyfour\ by
\citealt{suleimanov2011} and \oeight\ by \citealt{poutanen2014}, for
which the data selection and analysis do not yield self-consistent
results.)

\begin{figure}
\includegraphics[height=5in]{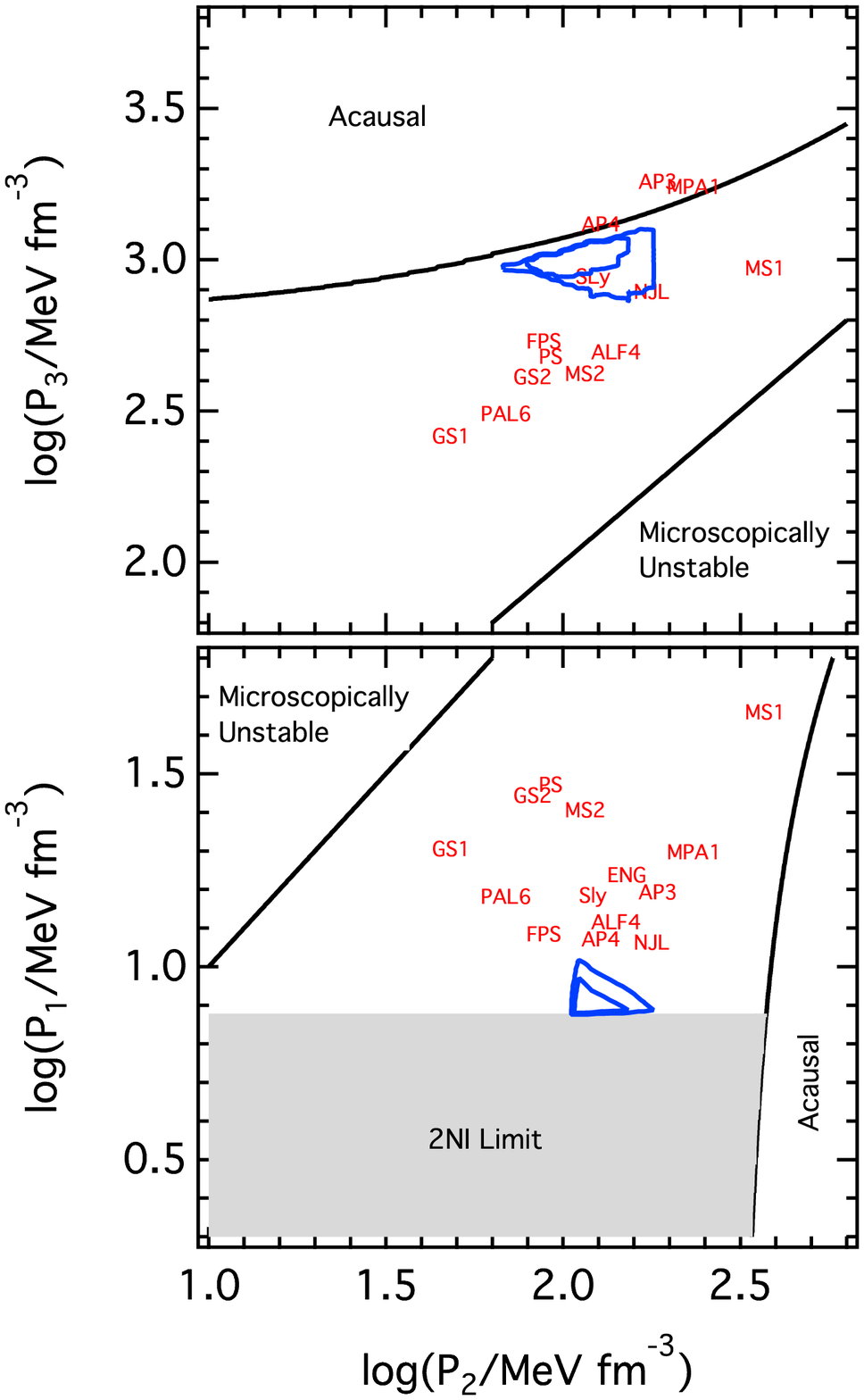}
\caption{The highest likelihood regions in the pressure of neutron
  star matter at 1.85~$\rho_{\rm ns}$ ($P_1$), 3.7~$\rho_{\rm ns}$
  ($P_2$), and 7.4~$\rho_{\rm ns}$ ($P_3$) obtained by a parametric
  inversion of all the neutron star radius measurements and requiring
  that it is consistent with the maximum mass requirement and
  low-energy experiment data. To visualize the structure of the
  three-dimensional posterior likelihood function, the contours
  outline the regions in which the posterior likelihoods drop down to
  $e^{-1/2}$ and $e^{-1}$ of the highest value. The allowed regions of
  the parameter space are consistent with the constraints from
  calculations based on low-energy scattering experiments, are
  microscopically stable, and ensure that the EoS remains causal.}
\label{fig:P1P2P3} 
\end{figure}


Figure~\ref{fig:P1P2P3} shows the posterior likelihoods over the
pressures at the three fiducial densities, as well as the microscopic
and experimental bounds on these pressures. The pressures at the
fiducial densities of a large number of proposed EoS are also shown.
The gray region labeled 2NI denotes the lower limit on $P_1$ coming
from the two-body interaction potential obtained at low
densities. $P_1$ is also constrained from above by the the relatively
small measured radii. Indeed, the values within the highest posterior
likelihood is lower than the pressure predicted by most proposed EoS
at that density.

\begin{figure}
\begin{minipage}{2.9in}
\includegraphics[height=2.8in, width=2.8in]{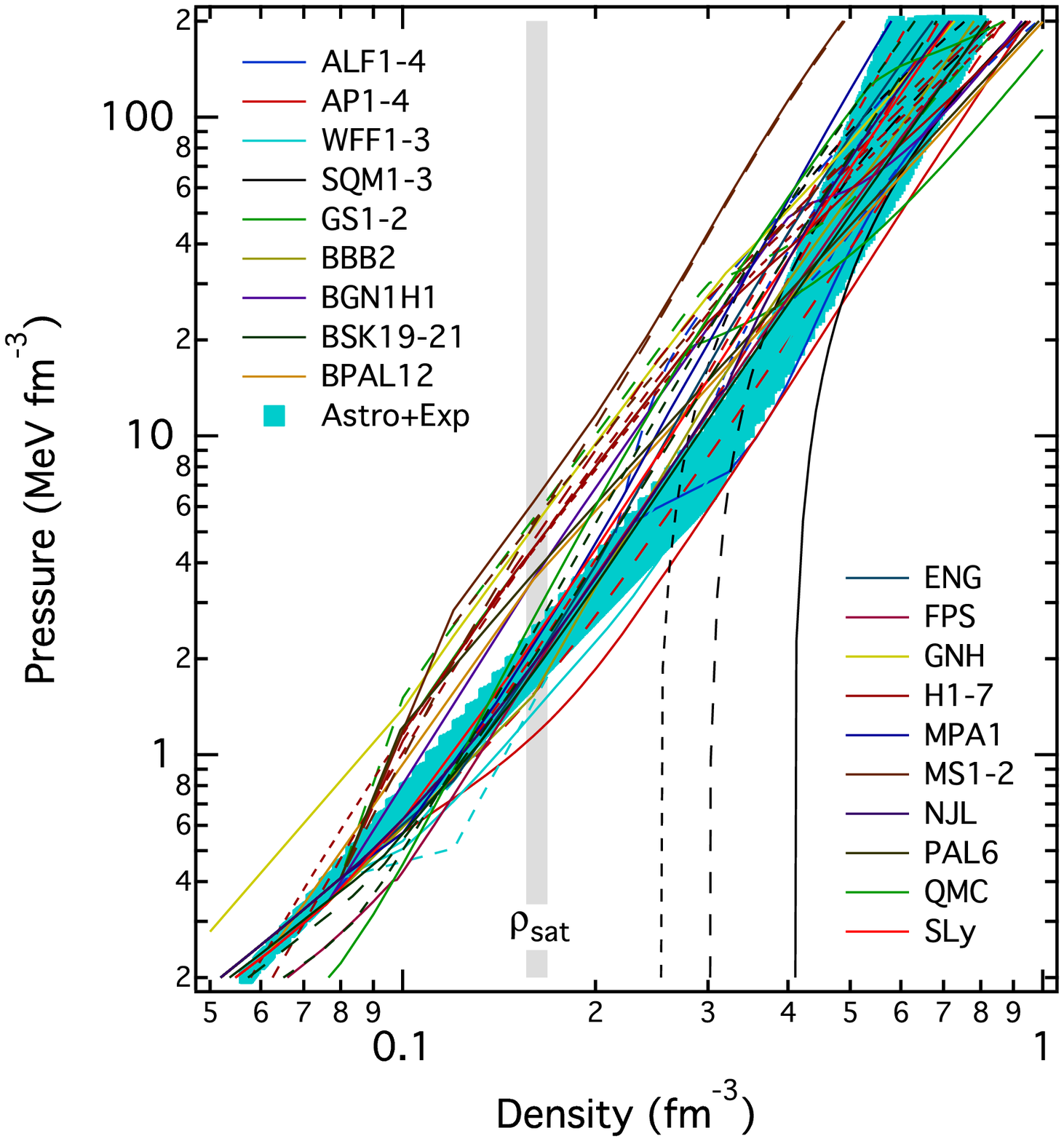}
\end{minipage}
\begin{minipage}{2.9in}
\includegraphics[height=2.8in, width=2.8in]{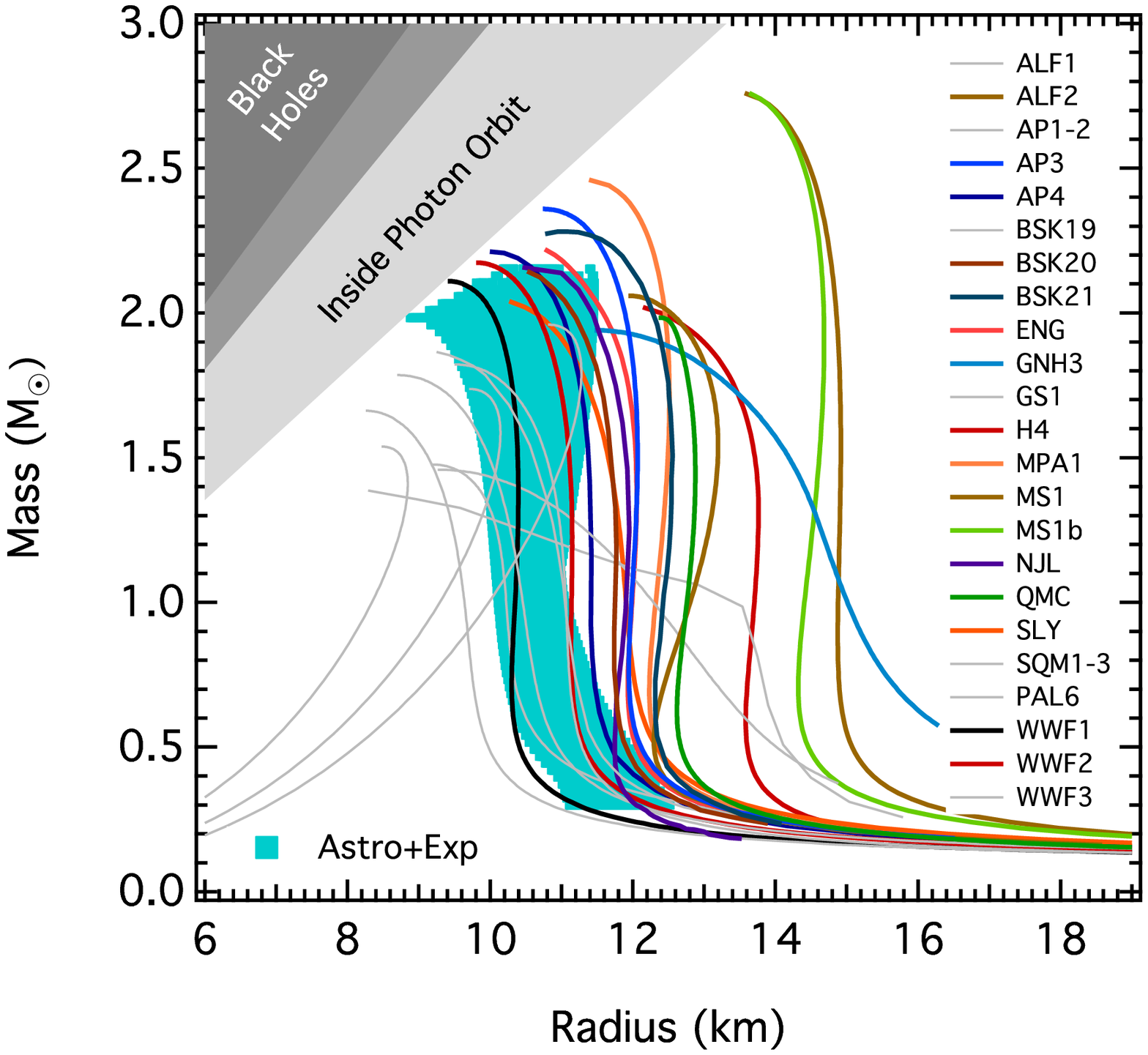}
\end{minipage}
\caption{The astrophysically inferred (left) EoS and (right)
  mass-radius relation corresponding to the most likely triplets of
  pressures that agree with all of the neutron star radius and low
  energy nucleon-nucleon scattering data and allow for a
  $M>1.97~M_\odot$ neutron star mass. The light blue bands show the
  range of pressures and the mass-radius relations that correspond to
  the region of the $(P_1, P_2, P_3)$ parameter space in which the
  likelihood is within $e^{-1}$ of its highest value. Around $1.5 \;
  M_\odot$, this inferred EoS predicts radii between $9.9-11.2$~km.}
\label{fig:eos_mr} 
\end{figure}

The combination of $P_2$ and $P_3$, on the other hand, is constrained
by the maximum mass requirement: a lower value of $P_2$ pushes $P_3$
to be as high as possible within the causality limit, whereas for
moderate to high values of $P_2$, which already lead to M-R relations
that allow high mass stars and are consistent with the radius
measurements, the allowed range of $P_3$ extends to lower values. The
combination of $P_2$ and $P_3$ exclude to high confidence the stiff
EoS such as MPA1 and MS1, which produce radii that are too large (see
also their inconsistency with $P_1$ in the lower panel). This
combination also excludes EoS with condensates, such as GS1, with
pressures that are too low to be consistent with the maximum mass
requirement.

Figure~\ref{fig:P1P2P3} shows that the combination of the radius
measurements with the low density experimental data and the
requirement of a $\sim 2\;M_\odot$ maximum mass pins down the
parameters of the EoS extremely well across a wide range of
supranuclear densities and points to a preferred EoS that is somewhat
softer than the nuclear EoS AP4 (a version of the APR EoS). To see
this on the mass-radius diagram, we also show in
Figure~\ref{fig:eos_mr} the mass-radius relation corresponding to the
most likely triplet of pressures as well as the range of mass-radius
relations for the region of the $(P_1, P_2, P_3)$ parameter space with
the highest likelihood.  We limit the range of masses in this figure
to $\le 2.2\;M_\odot$ because of the absence of any data to constrain
the relation at higher masses. As expected from the above discussion,
the preferred mass-radius relation lies to the left of most model
predictions and is closest to AP4, especially at low masses, where the
main uncertainty in AP4 is in the strength of the three-nucleon
interactions. It also rises along a nearly constant radius in order to
reach the $\sim 2~M\odot$ limit.  Depending on the choice of the
prior, the predicted radius for a $1.5 \; M_\odot$ neutron star is
between 9.9 and 11.2~km.

\section{Future Prospects}

\subsection{The future of radio pulsar mass measurements: New surveys, new binary systems, 
new measurements - and the SKA}
\label{sec:MSPnumbers}

\begin{figure}
\includegraphics[height=3in, width=3.5in]{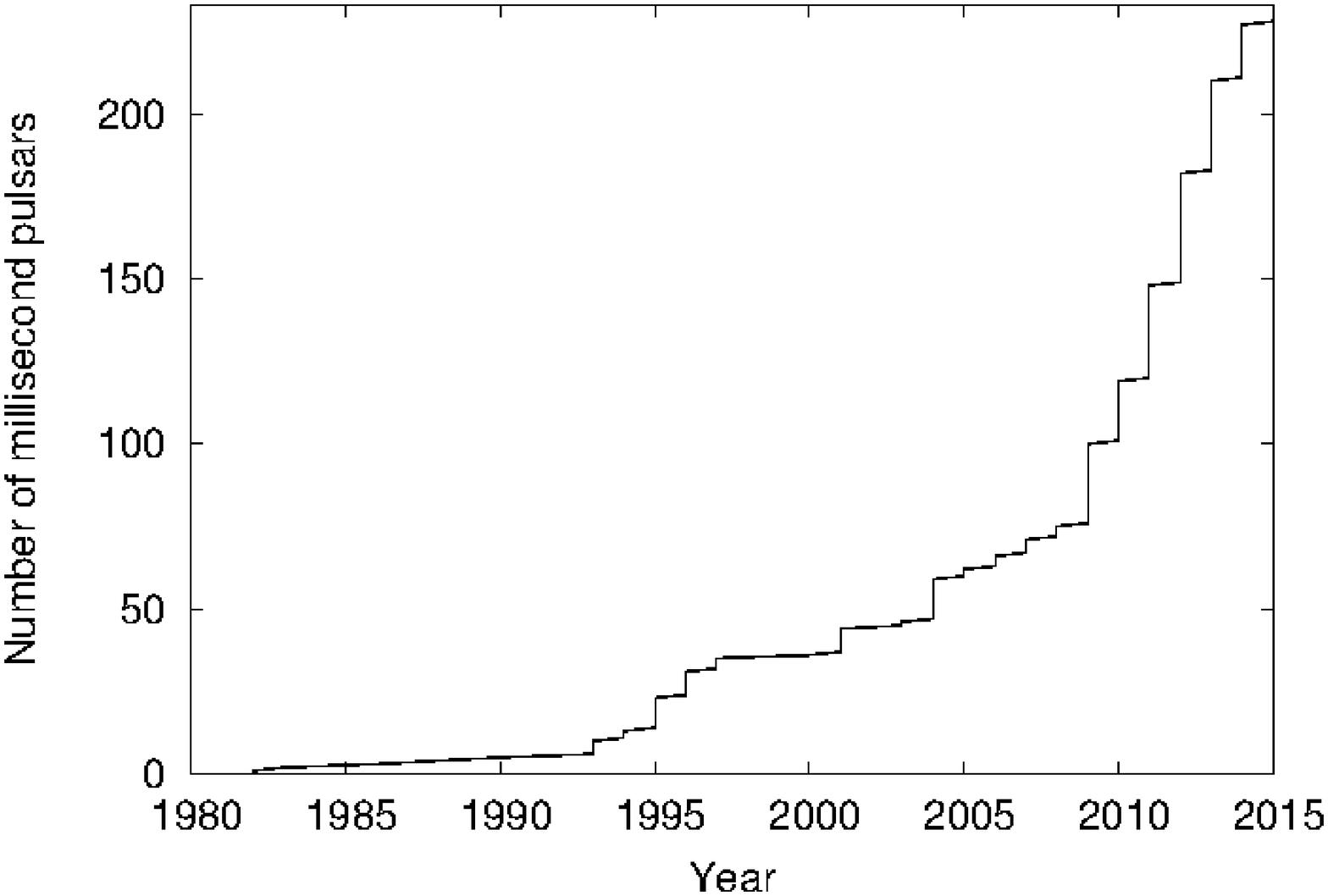}
\caption{Number of MSPs (with $P < 20$ ms) known in the Galactic disk
  as a function of time.  This figure does not take into account the
  more than 100 MSPs known in globular clusters. The large number of
  discoveries in recent years was enabled by improvements in the
  digital equipment being used in ground-based radio surveys and the
  discovery of many pulsar-like gamma-ray point sources by the Fermi
  satellite, which led to the discovery of many new radio MSPs. Data
  provided by Duncan Lorimer, see
  \texttt{http://astro.phys.wvu.edu/GalacticMSPs/GalacticMSPs.txt}.
}\label{fig_msp_numbers}
\end{figure}

We can expect the number of mass measurements to increase at an
accelerating rate over the next few years.  The reason for this is
that improvements in instrumentation, both in terms of new receivers
and the equipment used to detect and process the radio signals, are
revolutionizing the field of pulsar timing and searching.

More sensitive receivers with larger bandwidths and the advent of
broadband coherent dedispersion allow much more precise timing of the
known binary pulsars. This leads to more precise measurements of known
PK parameters and the detection of new ones, greatly increasing the
sample of NS mass measurements and the number of GR tests in different
systems. Furthermore, improved receiver sensitivities and larger
bandwidth, together with multi-beaming, increase the general
sensitivity to all sorts of pulsars, as in the case of the
phenomenally successful Parkes multibeam (PMB) surveys
\citep{manchester2001}. Since then, the newer generation of post-PMB
surveys like the new HTRU-South \citep{keith2010} and North surveys
\citep{barr2013}, the Green Bank drift scan survey
\citep{boyles2013,lynch2013}, the Green Bank North Celestial Cap
survey \citep{stovall2014}, the Arecibo ALFA survey
\citep{cordes2006,lazarus2013} and the Arecibo 327 MHz drift-scan
survey \citep{deneva2013} has been making full use of much improved
data handling capabilities and processing power to obtain a large
increase in spectral and time resolution (at a cost of 100 times the
data rate). This has paid off with a great increase in the sensitivity
to millisecond pulsars, particularly the distant ones, with a great
leap in their discovery rate. The number of discoveries per year is
shown in Fig.~\ref{fig_msp_numbers}; we see a 3-fold increase in the
number of MSPs since 2010. This will continue for the next few years,
as many of the new surveys are still starting. The rate of discovery
will further accelerate as soon as Square Kilometer Array (SKA) comes
online \citep{kramer2015}. 

Not all these new MSPs were found in blind surveys. More than 100 were
discovered in targeted radio searches of globular
clusters\footnote{See
  http://www.naic.edu/$\sim$pfreire/GCpsr.html}. This number is
expected to increase to thousands with the SKA \citep{hessels2015}. A
new and very successful search strategy -- with more than 60 new MSPs
discovered at the time of writing -- has been targeting unidentified
pulsar-like gamma-ray sources discovered with the {\it Fermi}
satellite (see, e.g., \citealt{ransom2011,keith2011,kerr2012}). Apart
from the intrinsic interest in discovering new MSPs, the gamma-ray
pulsars are important in the sense that they are biased {\em in favor}
of very short-period MSPs, which is the opposite of the outcome in
normal blind surveys. This contributes to a better understanding of
the underlying MSP population.

The large increase in the number of binary pulsars implies a large
increase in the number of NS mass measurements and tests of GR,
because discoveries of more edge-on systems will inevitably result
from a larger MSP sample, as well as discoveries of more triples,
eccentric MSP systems, and DNSs.  This trend is evident in
Tables~\ref{tab1} and \ref{tab2}, where most of the reported results
are relatively recent. Furthermore, the large sensitivity of planned
radio telescopes such as the SKA will enable much improved timing
precision of binary pulsars, new and old. This will further increase
the number and precision of neutron star mass measurements and
possibly allow a measurement of the upper mass limit for neutron
stars.

\subsection{The Future of Radius Measurements}

The increase in the number of neutron star radius measurements and the
continued reduction of the uncertainties will rely on a few factors.
Distance uncertainties are currently one of the bottlenecks in the
spectroscopic radius measurements. Therefore, one of the crucial
improvements in this technique can come from obtaining distance
measurements to many more neutron star X-ray binaries. The GAIA
mission and radio parallax measurements may provide distances to the
ones that are relatively nearby.

Radius measurements obtained by non-spectroscopic techniques, with
potential biases and systematic uncertainties that are different than
the spectroscopic ones will be another key factor in verifying the
current measurements. To this end, high signal-to-noise pulse profiles
of rotation- and accretion-powered X-ray pulsars will provide a major
step forward. Indeed, the technique of pulse profile modeling to
measure the neutron star compactness and radii defines the key
scientific objectives of several future X-ray missions, such as NASA's
upcoming NICER \citep{gendreau2012}, ISRO's Astrosat
\citep{agrawal2006} or the proposed LOFT \citep{feroci2012} missions.

Among these missions, NICER, a NASA Mission of Opportunity, is poised
to be launched to the International Space Station in 2016. NICER will
collect a large number of counts from a small number of
rotation-powered pulsars that emit thermally from their polar caps.
In combination with the refined theoretical models that have been 
developed to analyze these waveforms, these data will provide 
a new measurement of neutron star radii and equation of state. 

\subsection{Other Future Probes of Neutron Star Structure}

The double pulsar will be the first system that will allow, in a few
years, a measurement of the moment of inertia of a neutron star
\citep{lyne2004}. The moment of inertia, being a higher moment of the
mass distribution of the neutron star, is a powerful probe of its
internal structure and, hence, of its equation of state
\citep{morrison2004,lattimer2005}.

There are also significant prospects for radius measurements from
Advanced LIGO observations of coalescing neutron star binaries. The
characteristic frequencies of these waveforms can be used to obtain
information on the neutron star radius and deformability (see, e.g.,
\citealt{read2013}), which will shrink the current uncertainties in
the radius measurements and further constrain the dense matter
equation of state.

Finally, the equation of state of neutron stars can be inferred by
measuring the total luminosity and the spectrum of neutrinos emitted
during a core-collapse supernova explosion within our Galaxy.

\begin{issues}[FUTURE ISSUES]
\begin{enumerate}
\item Continued measurements of neutron star masses will make the
  inferred neutron star mass distributions more robust and help
  constrain the maximum neutron star mass by increasing the size of
  the sample.
\item Comparing current radius measurements against those made with
  other techniques will further reduce the uncertainties in the neutron
  star radii.
\item Multi-messenger probes such as gravitational waves, neutrinos,
  and moment-of-inertia measurements will open new avenues into neutron 
  star structure. 
\item Advances in lattice QCD will provide a first-principles
  calculation of the composition and interactions of the dense matter
  equation of state, which can be compared with the results from
  neutron stars and laboratory experiments. 
\item The hot dense matter equation of state that is relevant to
  supernovae and neutron star mergers requires further study.
\end{enumerate}
\end{issues}

\section*{DISCLOSURE STATEMENT}
The authors are not aware of any affiliations, memberships, funding, or 
financial holdings that might be perceived as 
affecting the objectivity of this review.

\section*{ACKNOWLEDGMENTS}
We thank John Antoniadis, Ben Lackey, Jim Lattimer, Michael Kramer,
Duncan Lorimer, Dimitrios Psaltis, Jocelyn Read, and Mallory Roberts
for graciously providing materials for this review. We thank Gordon
Baym, Tolga G\"uver, Ramesh Narayan, and Dimitrios Psaltis for numerous
useful conversations and comments on the manuscript. F.\"O. gratefully
acknowledges support from NSF grant AST 1108753. P.C.C.F. gratefully
acknowledges financial support by the European Research Council for
the ERC Starting grant BEACON under contract No. 279702.

\bibliography{NS_Mass_Radius_ARAA}

\end{document}